\documentclass[sigconf]{acmart} 

\usepackage{subcaption}
\usepackage{graphicx}
\usepackage{lipsum}
\usepackage{makecell}
\usepackage[noabbrev,capitalise]{cleveref}
\usepackage{multirow}

\usepackage[title]{appendix}
\usepackage{caption}

\AtBeginDocument{%
  }

\copyrightyear{2024}
\acmYear{2024}
\setcopyright{rightsretained}
\acmConference[RecSys Challenge '24]{ACM RecSys Challenge 2024}{October 14--18, 2024}{Bari, Italy}
\acmBooktitle{ACM RecSys Challenge 2024 (RecSys Challenge '24), October 14--18, 2024, Bari, Italy}
\acmDOI{10.1145/3687151.3687152}
\acmISBN{979-8-4007-1127-5/24/10}

\begin{document}

\title{
    EB-NeRD: A Large-Scale Dataset for News Recommendation
} 

\author{Johannes Kruse}
\email{johannes.kruse@jppol.dk}
\orcid{0009-0007-5830-0611}
\affiliation{
  \institution{JP/Politikens Media Group}
  \city{Copenhagen}
  \country{Denmark}
}
\additionalaffiliation{
  \institution{Technical University of Denmark}
  \city{Kongens Lyngby}
  \country{Denmark}
}

\author{Kasper Lindskow}
\email{kasper.lindskow@jppol.dk}
\orcid{0009-0004-6412-0930}
\affiliation{
  \institution{JP/Politikens Media Group}
  \city{Copenhagen}
  \country{Denmark}
}
\additionalaffiliation{
  \institution{Copenhagen Business School}
  \city{Frederiksberg}
  \country{Denmark}
}

\author{Saikishore Kalloori}
\email{ssaikishore@ethz.ch}
\orcid{0000-0002-9669-0602}
\affiliation{
  \institution{ETH Zürich}
  \city{Zürich}
  \country{Switzerland}
}

\author{Marco Polignano}
\email{marco.polignano@uniba.it}
\orcid{0000-0002-3939-0136}
\affiliation{
  \institution{University of Bari Aldo Moro}
  \city{Bari}
  \country{Italy}
}

\author{Claudio Pomo}
\email{claudio.pomo@poliba.it}
\orcid{0000-0001-5206-3909}
\affiliation{
  \institution{Politecnico di Bari}
  \city{Bari}
  \country{Italy}
}

\author{Abhishek Srivastava}
\email{abhishek@iimv.ac.in}
\orcid{0000-0002-7113-6947}
\affiliation{
  \institution{Indian Institute of Management Visakhapatnam}
  \city{Visakhapatnam}
  \country{India}
}

\author{Anshuk Uppal}
\email{ansup@dtu.dk}
\orcid{0009-0008-2427-5857}
\affiliation{
  \institution{Technical University of Denmark}
  \city{Kongens Lyngby}
  \country{Denmark}
}

\author{Michael Riis Andersen}
\email{miri@dtu.dk}
\orcid{0000-0002-7411-5842}
\affiliation{
  \institution{Technical University of Denmark}
  \city{Kongens Lyngby}
  \country{Denmark}
}

\author{Jes Frellsen}
\email{jefr@dtu.dk}
\orcid{0000-0001-9224-1271}
\affiliation{
  \institution{Technical University of Denmark}
  \city{Kongens Lyngby}
  \country{Denmark}
}

\renewcommand{\shortauthors}{Kruse et al.}
\acmArticleType{Review}



\begin{abstract}
    Personalized content recommendations have been pivotal to the content experience in digital media from video streaming to social networks. However, several domain specific challenges have held back adoption of recommender systems in news publishing. To address these challenges, we introduce the Ekstra Bladet News Recommendation Dataset (EB-NeRD). The dataset encompasses data from over a million unique users and more than $37$ million impression logs from Ekstra Bladet. It also includes a collection of over $125{,}000$ Danish news articles, complete with titles, abstracts, bodies, and metadata, such as categories.  
    EB-NeRD served as the benchmark dataset for the RecSys '24 Challenge, where it was demonstrated how the dataset can be used to address both technical and normative challenges in designing effective and responsible recommender systems for news publishing.
    The dataset is available at: \url{https://recsys.eb.dk}.
\end{abstract}

\maketitle

\section{Introduction}
\label{sec:introduction}
Recommender systems (RS) have become an integral component of the web and are central to producing massive commercial value gains in e-commerce (e.g., Amazon), social media (e.g., Meta), entertainment media (e.g., Netflix and Spotify), and online advertising (e.g., Google and Criteo) \citep{Jannach2019}. However, even though recommender systems have also been adopted by news publishers \citep{Radecka2019}, several domain-specific 1) technical and 2) normative challenges have held back adoption at scale in the news publishing domain.

\citet{Wu2023} recently highlighted several technical challenges (1) in the news recommendation scenario in a call for further research on RS in news. 
Firstly, news articles are published in a continuous flow, and they tend to expire quickly, resulting in a severe cold-start problem \citep{Das2007}. Classic recommender systems, such as collaborative filtering (CF) \citep{Koren2008} or factorization machines (FM) \citep{Rendle2010}, which have been successfully adapted in other fields, are therefore not very applicable.
Secondly, explicit user ratings for news articles are rarely found on news platforms, necessitating modeling users' ever-changing news interests based on implicit feedback alone, i.e., based on their browsing behavior \citep{Ilija2013}.
Thirdly, effective RS for news must leverage textual information from news articles \citep{Kompan2010}.

In recent years, attempts have been made to address these challenges by developing and testing recommender systems designed specifically for the news domain \citep{Karimi2018, raza2020, Feng2020}. 
Most significantly, \citet{Wu2020MIND} identified recommender systems that effectively addressed some of the challenges while demonstrating strong performance on the MIcrosoft News Dataset (MIND), a dataset containing news consumption data from the major news aggregator MSN News\footnote{\url{https://www.msn.com}}.
However, most news recommendation datasets are not public and are tailored to specific use cases, possibly limiting the applicability of the derived models when introduced to different contexts or broader applications. For instance, it is unclear whether the performance achieved by \citet{Wu2020MIND} can be generalized to classical digital news publishers, which have news consumption and content profiles that differ markedly from news aggregators (see Section \ref{sec:existing_datasets}).

The normative challenges (2) stem from the fact that news recommenders, when implemented at scale, perform a deeply editorial function that affects the editorial profile of a news publisher. This creates a need to align the output of news recommender systems with editorial values \citep{lu2020_editoral_values} and social values \citep{helberger2019}. 
Typically, attempts to do so have taken traditional beyond-accuracy metrics as the point of departure \citep{raza2020} and sought to connect them to editorial values at the level of a news publisher \citep{lu2020_editoral_values} or societal values associated with different normative theories of news publishers’ role in democracy \citep{Vrijenhoek2021_mission, Vrijenhoek2021_radio}. However, research on beyond-accuracy metrics relevant to the news domain as well as ways to integrate beyond-accuracy objectives in the optimization targets of recommender systems is still nascent, as we argue below.

\paragraph{Contributions} 
To enable and encourage research on both the technical and normative challenges associated with news recommendations, we make the following contributions:
\begin{itemize}
    \item \textbf{EB-NeRD Dataset}: We present the Ekstra Bladet News Recommendation Dataset (EB-NeRD), a rich dataset collected from the user behavior logs of \emph{Ekstra Bladet}, a classical legacy Danish newspaper published by JP/Politikens Media Group in Copenhagen, Denmark.
    \item \textbf{Descriptive Statistics}: We provide detailed descriptive statistics of the dataset to facilitate understanding and effective utilization.
    \item \textbf{Beyond-Accuracy Evaluation Framework}: We present an extensive, albeit preliminary, framework for beyond-accuracy evaluation, enabling a more comprehensive assessment of news recommendation systems.
\end{itemize}

\section{Related Work}
\label{sec:related_work}
\subsection{News Recommendations}
News recommendation systems are designed to enhance the news reading experience by alleviating information overload and presenting users with a personalized selection of articles from a vast catalog. Typically, this involves ranking a set of news articles, ordering them from most to least likely to be clicked by the user.
To achieve this, various sources of data can be leveraged, including the user’s click history, session details (such as time and device), user metadata (such as age), and the content of the news articles. 
Research in news recommendation has primarily focused on addressing two key challenges: 1) representing the raw text found in news articles, and 
2) accounting for the dynamic nature of the the user's news interest \citep{Tian2021, an2019-lstur, wu2019-npa, wu2019-nrms}.

Collaborative filtering has been less commonly applied in news recommendation due to the severe cold-start problem -- by the time sufficient behavioral data is collected from users, the relevance of news articles has already decayed \citep{raza2020, Wu2020MIND}.
Consequently, early research in news recommendation predominantly focused on content-based methods which aspire to generate personalized user profiles that are based on the metadata extracted from the latest news they have read; for example, using keywords \citep{Oh2014}, topic modeling \citep{Wang2018}, or semantic features \citep{Huang2013} extracted from the news articles. 
More recently, news recommendation approaches have increasingly adopted deep learning techniques, where representations of both news articles and user interests are learned end-to-end \citep{acharya2023, era_llm_2024, Wu2021-newsBert, Qi2021-know-interactive}. 
These advancements in deep learning have blurred the traditional distinctions between content-based, collaborative filtering, and hybrid systems, as many contemporary methods now integrate elements from multiple approaches. 
In response, \citet{Wu2023} propose classifying recommender systems based on the core challenges they address, rather than relying solely on traditional categories.

\subsection{Existing Datasets}
\label{sec:existing_datasets}
Personalized news recommendation presents both intriguing research challenges and significant practical benefits. However, most research in this area is conducted on proprietary datasets that are not publicly available, such as MSN News \citep{wu2019-npa, an2019-lstur, wu2022big_industry}, Sina News \citep{Gu2016}, and Twitter \citep{Abel2011}. In contrast, only a few publicly available news recommendation datasets exist \citep{Gulla2017_adressa, Kille-plista, Moreira2018_globo, Wu2020MIND}.

Currently, the datasets most akin to EB-NeRD are Adressa \citep{Gulla2017_adressa} and MIND \citep{Wu2020MIND}. A summary of these datasets can be found in \cref{tab:existing_datasets}.
The Microsoft News Dataset (MIND) was released by \citet{Wu2020MIND}. Microsoft, a major news aggregator, sourced content from a multitude of news publishers for this dataset. It showcases interactions from a 1 million users across $161{,}013$ English news articles, accumulating a total of $24{,}155{,}470$ clicks. 
The Adressa dataset, put forth by \citet{Gulla2017_adressa}, offers insights from the Adresseavisen website, a Norwegian digital news publisher. Over a span of ten weeks, this dataset contains interactions of $3.08$ million users with $48{,}486$ Norwegian news articles, and $27.22$ million clicks. 

While all of the datasets offer a range of features, there are clear distinctions between them. 
Each dataset provides essential news-related information, capturing elements like the title, category, and entities. Although MIND includes the body feature, it is not available in the public version of the dataset. Instead, users must scrape the body content. Even though \citet{Wu2020MIND} provides a helpful script for this task, some URLs have expired and are no longer accessible, raising concerns about the long-term availability of this feature.
In contrast, EB-NeRD distinguishes itself by offering the title, abstract, and body of articles directly in the public version of the dataset. 

In terms of non-textual features, MIND is notably limited, providing only user-ID, article-ID, and event-time.
Adressa and EB-NeRD, however, present a richer set of features, including read-time, location, and even the user's subscription status. Another limitation of MIND is that it only tracks front-page events, while both Adressa and EB-NeRD capture user activity more comprehensively, including interactions on both the front page and on article pages.
Additionally, the origin of the data further differentiates these datasets. MIND gathers its data from a news aggregator, whereas both EB-NeRD and Adressa are based on data sourced directly from news publishers.
Given the unique patterns exhibited by different news providers and the cultural nuances that influence news consumption across countries, we believe that a high-quality, low-resource language news recommendation dataset like EB-NeRD holds significant value for the news recommendation community. For instance, it remains unclear whether insights derived from news aggregators are directly applicable to news publishers.
\begin{table*}[tbp]
  \caption{Comparison of EB-NeRD Dataset with existing public news recommendation datasets.}
  \label{tab:existing_datasets}
  \centering
  \begin{tabular}{llllllll}
    \toprule
    Name    
        & Type
        & Language
        & \# Users 
        & \# News 
        & \# Clicks 
        & News information  
        & Features 
        \\
    \midrule
    MIND    
        & Aggregator    
        & English     
        & $1{,}000{,}000$
        & $161{,}013$
        & $24{,}155{,}470$
        & \makecell[l]{
            title, abstract, category, \\ 
            subcategories, entities,\\
            URL}
        & \makecell[l]{
            user-id, article-id, event-time\\ 
            }
        \\
    \midrule
    Adressa 
        & Publisher     
        & Norwegian     
        & $3{,}083{,}438$
        & $48{,}486$
        & $27{,}223{,}576$
        & \makecell[l]{
            title, body, category,\\
            entities, URL
            }
        & \makecell[l]{
            user-id, article-id, event-time,\\
            read-time, publish-time,\\
            session-boundary, location,\\
            subscription-status, \\
            device-type, OS
            }
        %
        \\
    \midrule
    EB-NeRD 
        & Publisher     
        & Danish  
        & $1{,}103{,}602$
        & $125{,}541$
        & $37{,}966{,}985$
        & \makecell[l]{
            title, abstract, body, \\
            category, subcategories,\\
            entities, URL, sentiment, \\
            topics            
            }
        & \makecell[l]{
            user-id, article-id, event-time, \\
            read-time, publish-time,\\
            session-boundary, location,\\
            subscription-status, \\
            device-type, age, gender, \\ 
            scroll-percentage}
        \\
    \bottomrule
  \end{tabular}
\end{table*}

\begin{figure}[tbp]
    \centering
    \begin{subfigure}{0.30\textwidth} 
        \centering
        \includegraphics[width=1\textwidth]{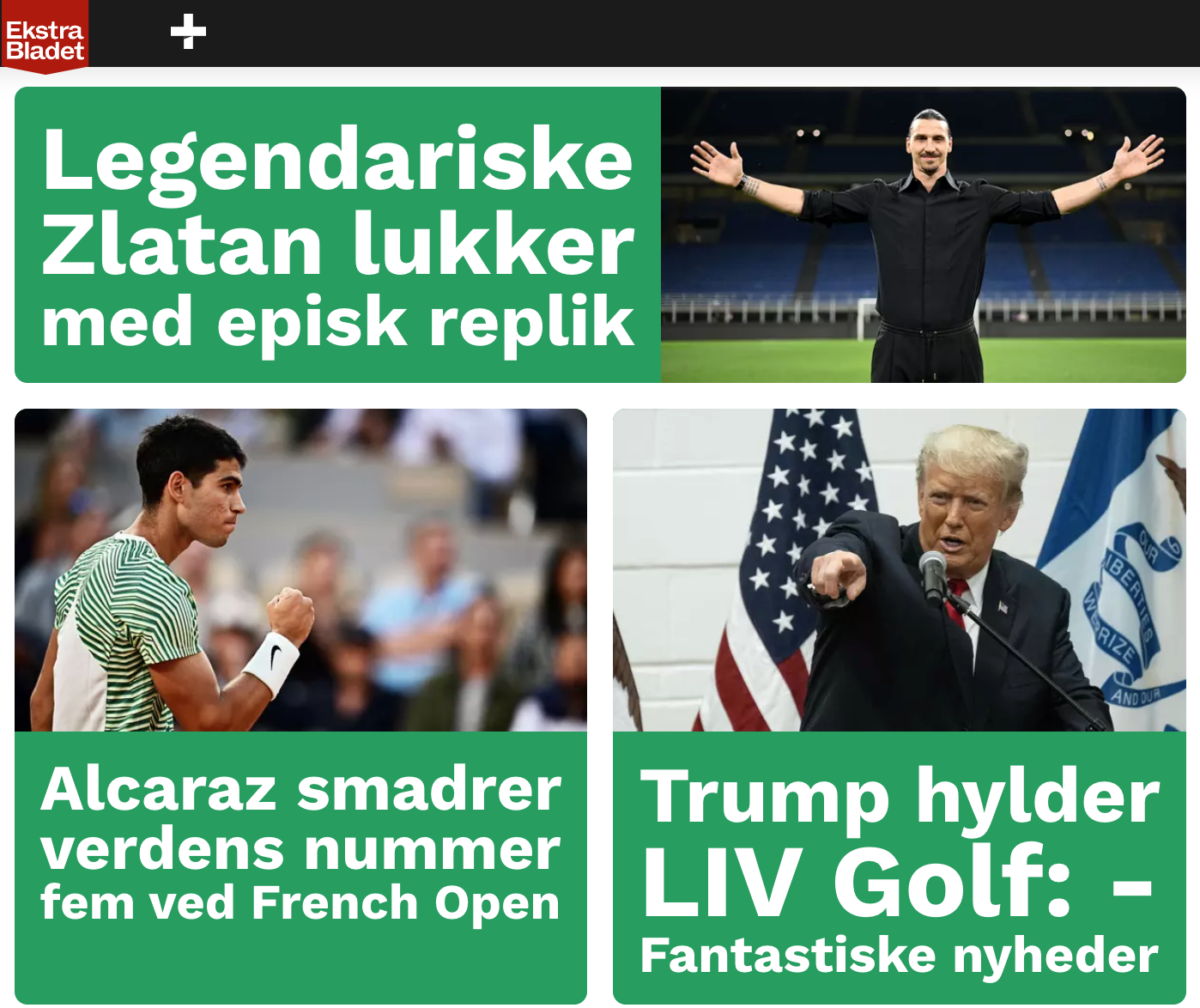}
        \caption{An example Ekstra Bladet front page}
        \label{fig:eb_front_page_example}
    \end{subfigure}
    \hfill
    \begin{subfigure}{0.4\textwidth} 
        \centering
        \includegraphics[width=1\textwidth]{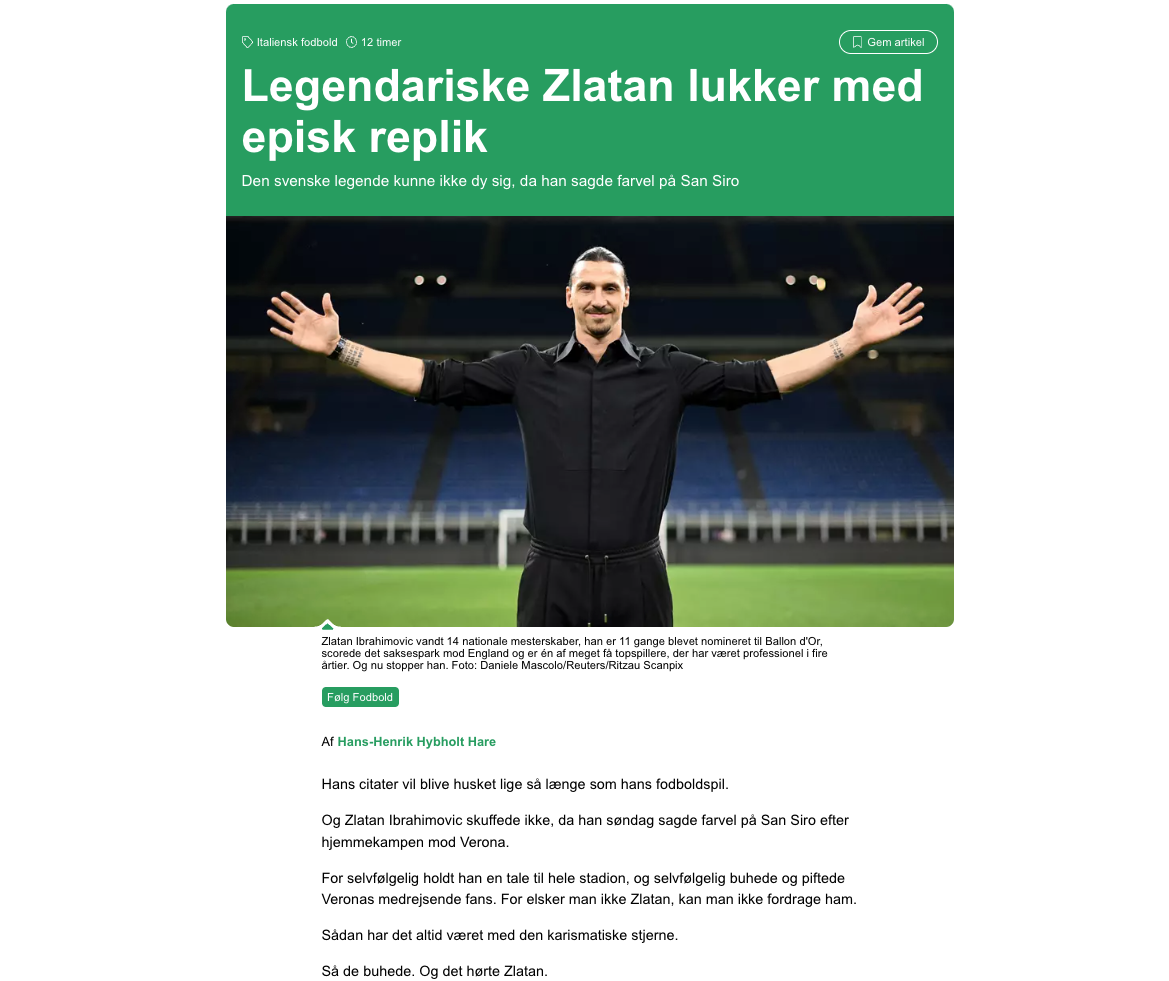}
        \caption{An example Ekstra Bladet article-page}
        \label{fig:eb_article_page_example}
    \end{subfigure}
    \caption{Examples of a front page and an article page from the Ekstra Bladet website.}
    \Description[Front page and article page examples]{
        This figure presents two examples from the Ekstra Bladet website. 
        (a) shows an example of a front page, featuring a collection of headlines, images, and a logo at the top. It represents the layout typically seen on the main page of the website, highlighting various news stories and sections.
        (b) shows an example of an article page, focusing on a specific news story. The page includes a large image of Zlatan Ibrahimović along with the article's text and related media, showcasing the detailed format used for individual stories on the website.
    }
    \label{fig:eb_front_article_page}
\end{figure}

\subsection{Dataset Description}
The Ekstra Bladet News Recommendation Dataset (EB-NeRD) is a large-scale Danish dataset created by Ekstra Bladet to support advancements and benchmarking in news recommendation research.

\subsubsection{Dataset Construction}
\label{sec:dataset-construction}
The dataset was collected from the impression logs of active users at Ekstra Bladet\footnote{\url{https://ekstrabladet.dk}} over a 6-week period from April 27 to June 8, 2023.
Active users were defined as those who had at least 5, but no more than 1,000, news clicks during a 3-week period from May 18 to June 8, 2023.
The lower threshold ensures that users had a minimum level of engagement, while the upper threshold helps exclude abnormal activity. 
To protect user privacy, each user was decoupled from the production system and securely hashed into an anonymized ID using one-time salt hashing \citep{joyofcryptography}.

Each impression log records the news articles viewed by a user during a visit at a specific time, which have been shuffled to remove positional bias, along with the articles that were clicked. The in view 
It includes details such as whether the user was on the front page or an article page, the timestamp, time spent, scroll percentage, the device used for browsing (i.e., computer, tablet, or mobile), and the user's subscription status.
Additionally, users with accounts may have provided personal details such as their gender, postal code, and age. 

Along with the impression logs, the dataset includes a collection of Danish news articles featuring their titles, abstracts, bodies, and metadata, including categories.
To support further research, such as knowledge-aware news recommendation, we have enriched the dataset with entities, topics, and sentiment labels extracted from the titles, abstracts, and bodies of the news articles using proprietary models. For example, in the article title shown in \cref{fig:eb_front_page_example}, \textit{``Legendariske Zlatan lukker med episk replik''}, \textit{``Zlatan''} is recognized as a person entity.

The EB-NeRD dataset includes training, validation, and test splits. Each data split consists of three files (see \cref{appendix:dataset-overview}): 
1) impression logs for the 7-day period (\cref{tab:behaviors_parquet}), 
2) users' click histories (\cref{tab:history_parquet}), which contain 21 days of clicked news articles prior to the data split’s impression logs, and 
3) all articles available during the period (\cref{tab:articles_parquet}).
Additionally, for evaluating beyond-accuracy metrics, a hidden test set includes $200{,}000$ synthesized impressions. The articles shown for these impressions are drawn from the same set of $250$ newly published articles available at the start of the test set period. It is important to note that these impressions are used solely for the beyond-accuracy evaluation.

\subsubsection{Dataset Statistics}
\label{sec:dataset-analysis}
\begin{figure}[tbp]
    \centering
    \begin{subfigure}{0.23\textwidth}
        \centering
        \includegraphics[width=1.0\textwidth]{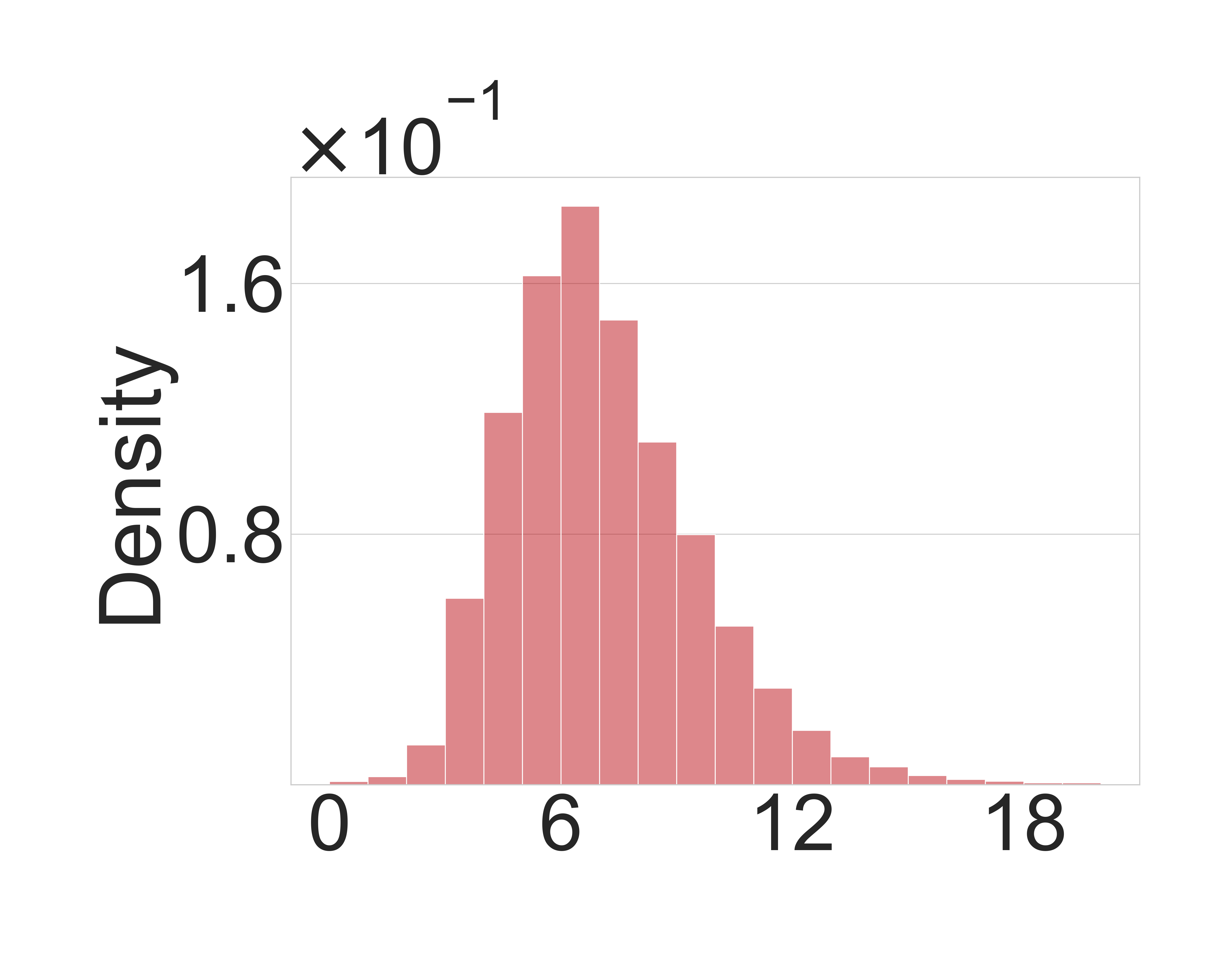}
        \caption{Title length}
        \label{fig:title_len}
    \end{subfigure}
    \begin{subfigure}{0.23\textwidth}
        \centering
        \includegraphics[width=1.0\textwidth]{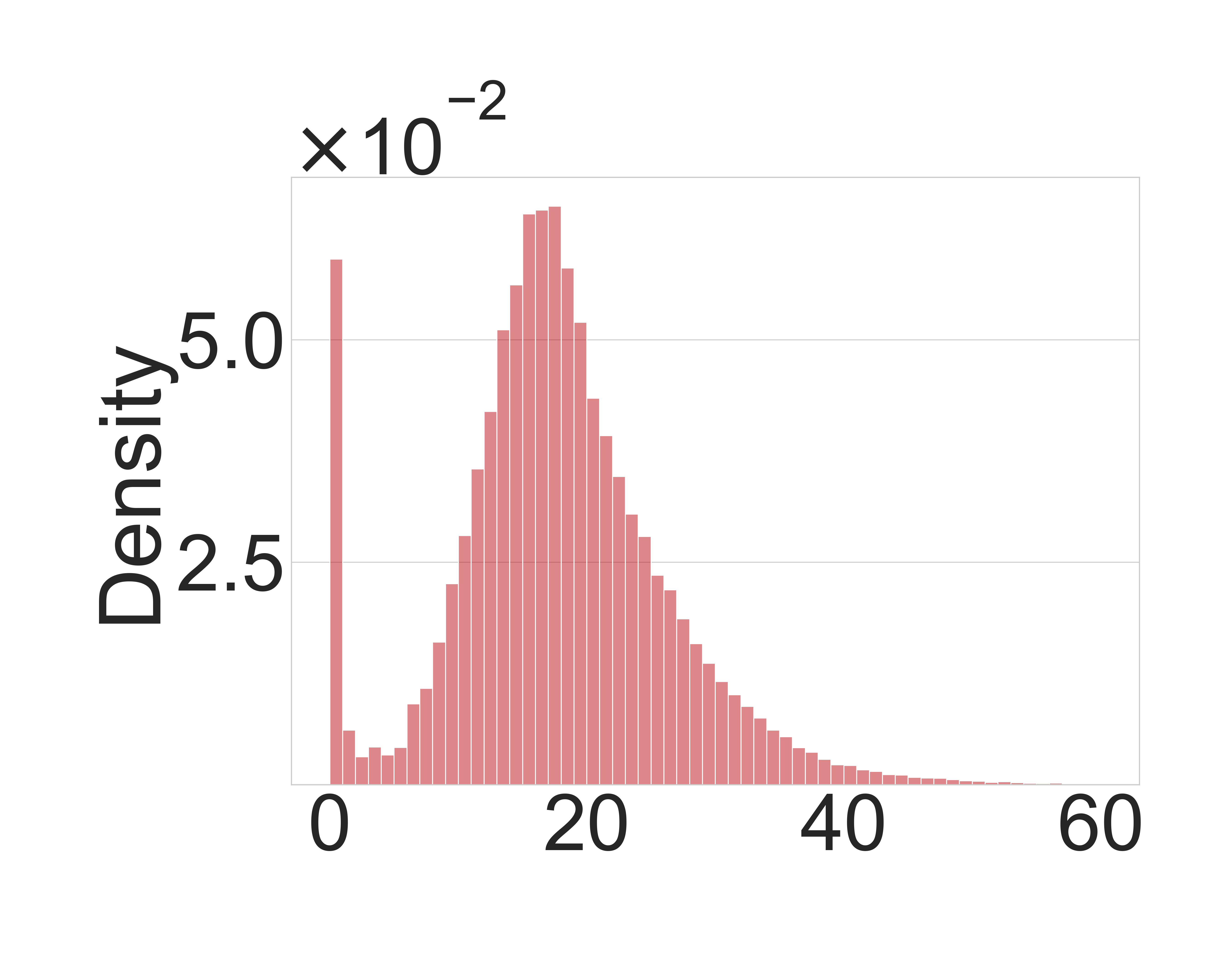}
        \caption{Abstract length}
        \label{fig:subtitle_len}
    \end{subfigure}
    \begin{subfigure}{0.23\textwidth}
        \centering
        \includegraphics[width=1.0\textwidth]{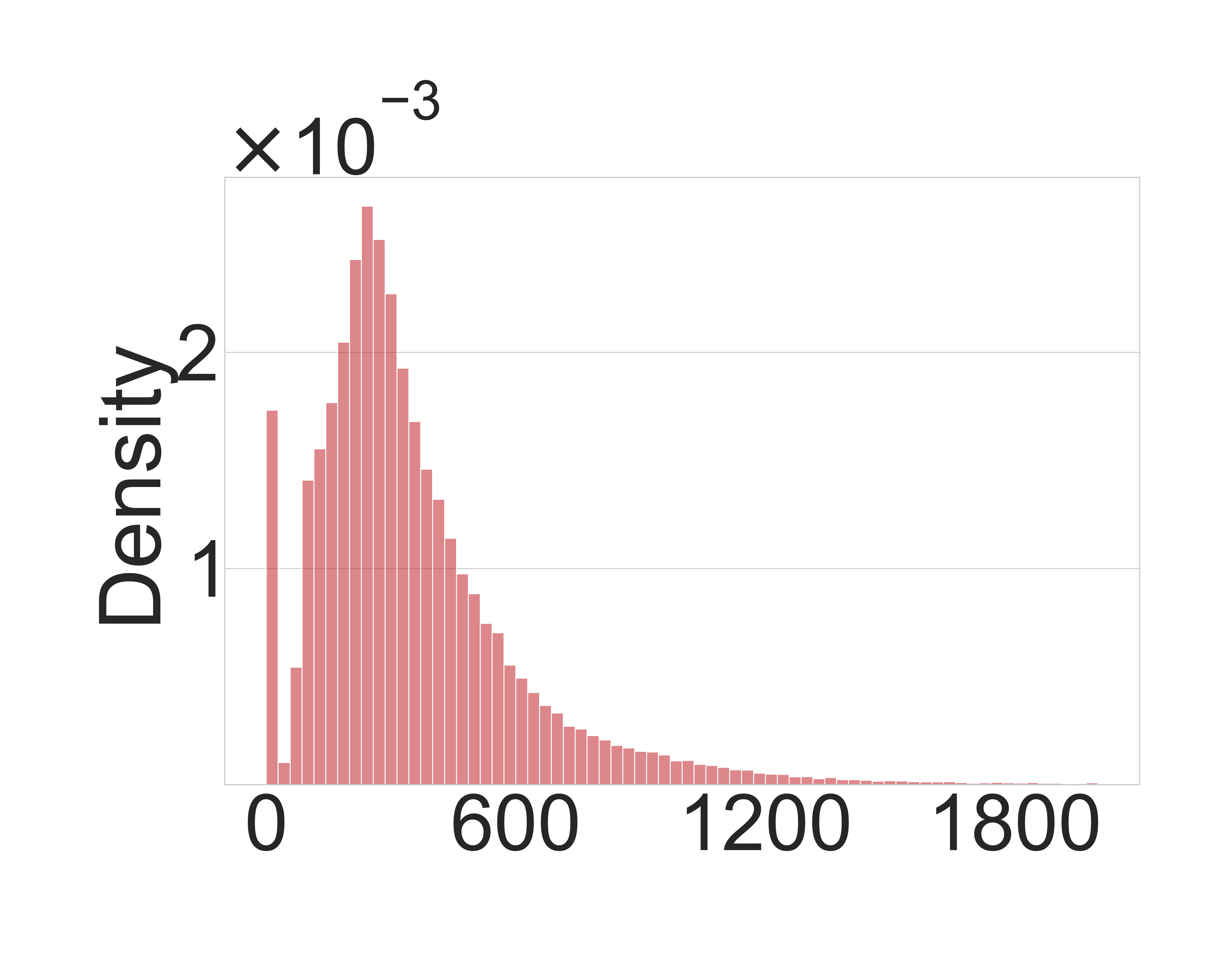}
        \caption{Body length}
        \label{fig:body_len}
    \end{subfigure}
    \begin{subfigure}{0.23\textwidth}
        \centering
        \includegraphics[width=1.0\textwidth]{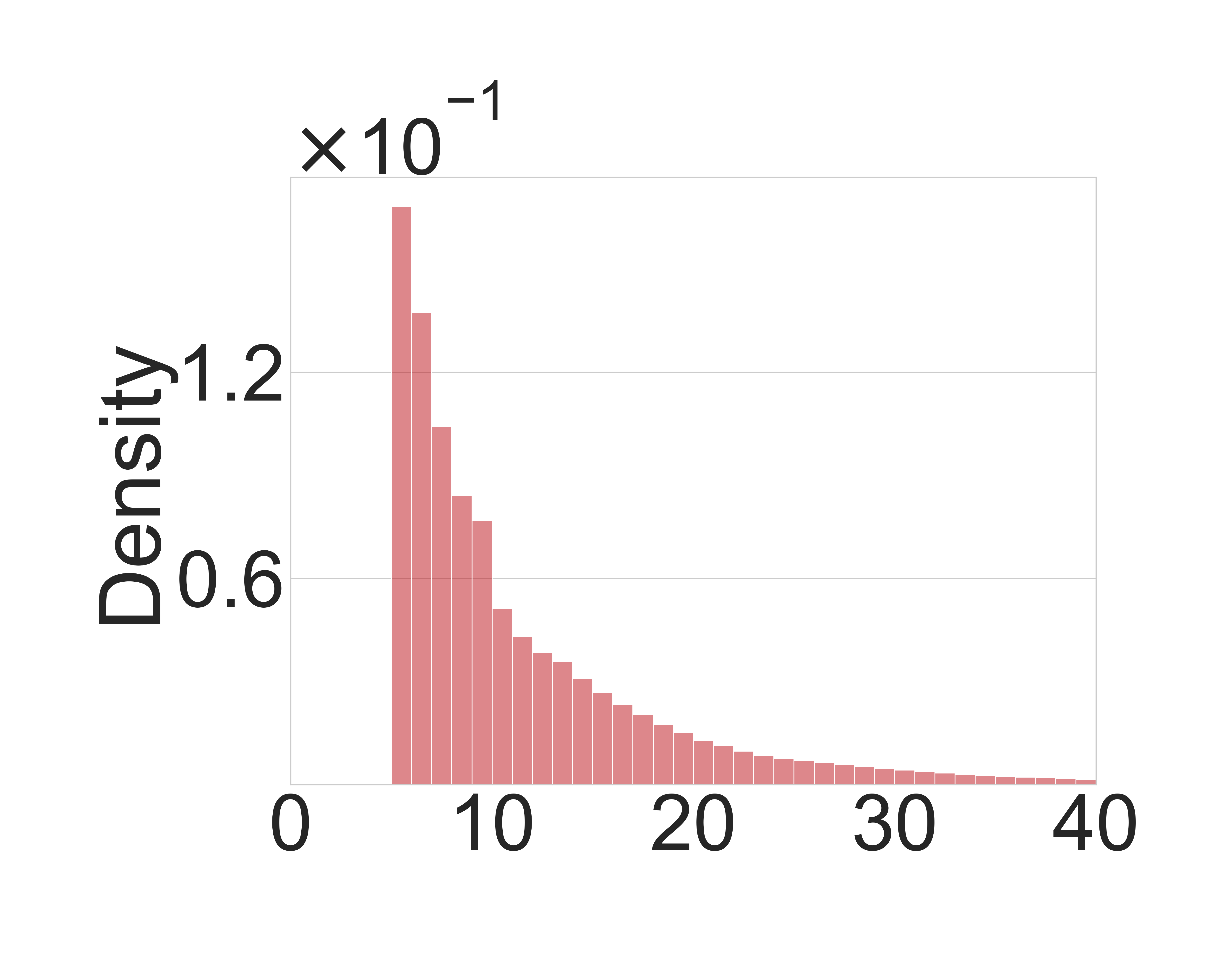}
        \caption{In view length}
        \label{fig:inview_len}
    \end{subfigure}
    \begin{subfigure}{0.23\textwidth}
        \centering
        \includegraphics[width=1.0\textwidth]{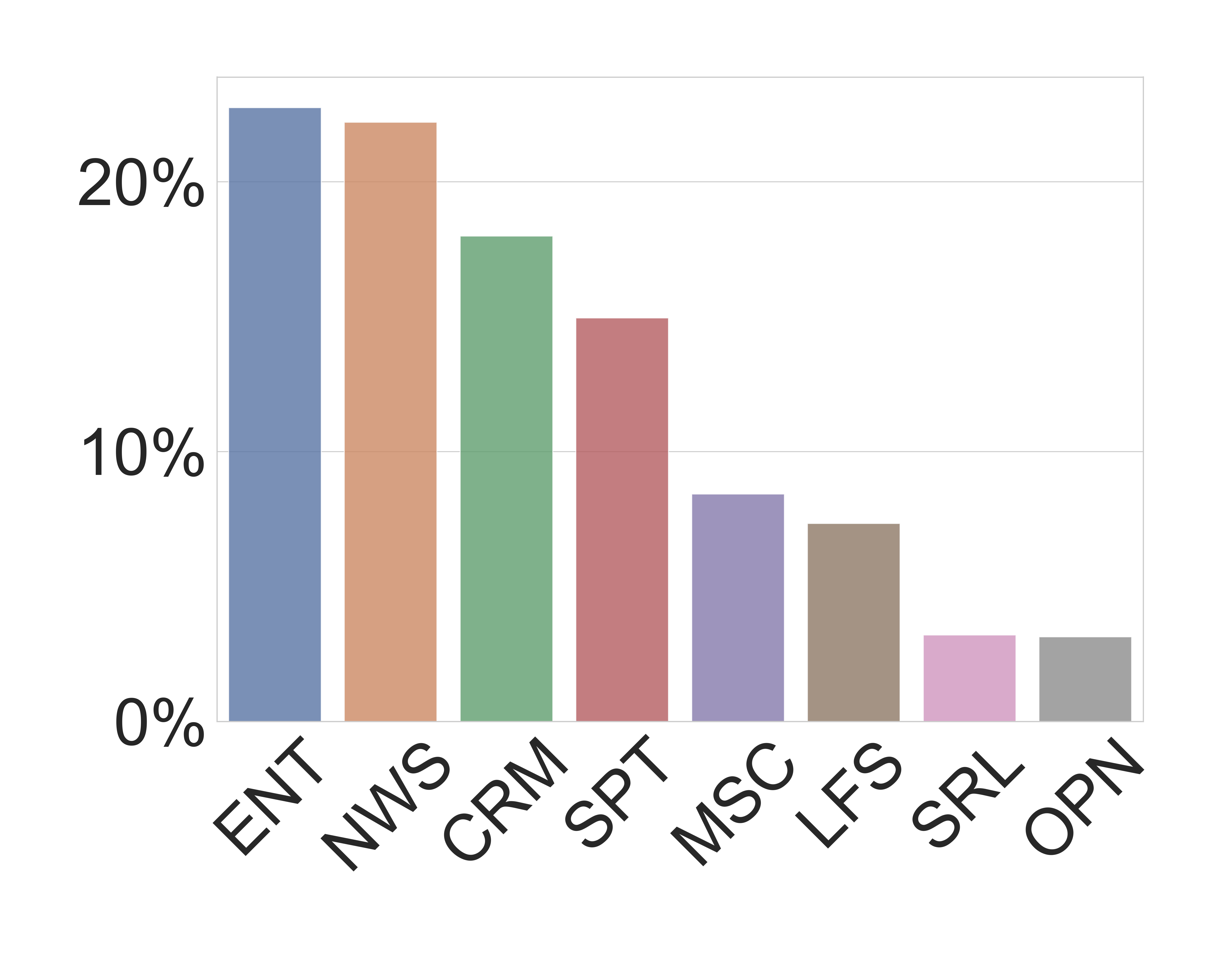}
        \caption{Article category distribution.}
        \label{fig:category_distribution}
    \end{subfigure}
    \caption{
        Key statistics of EB-NeRD. 
        The categories in (e) include entertainment (ENT), news (NWS), crime (CRM), sports (SPT), miscellaneous (MSC), lifestyle (LFS), sex and relationships (SRL), and opinion (OPN).
    }
    \Description[Key statistics of EB-NeRD.]{
        This figure presents key statistics of the EB-NeRD dataset across five subfigures. 
        (a) shows a bar plot representing the distribution of title lengths for articles, 
        (b) shows a bar plot representing the distribution of abstract lengths for articles, 
        (c) shows a bar plot representing the distribution of body lengths for articles, 
        (d) shows the distribution of the in-view content length, which refers to the portion of the article visible without scrolling.
        (e) presents the distribution of articles across different categories, including entertainment (ENT), news (NWS), crime (CRM), sports (SPT), miscellaneous (MSC), lifestyle (LFS), sex and relationships (SRL), and opinion (OPN). 
    }
    \label{fig:data_analysis}
\end{figure}

A overview of EB-NeRD's key statistics can be found in \cref{tab:EB-dataset-overview-table} and \cref{fig:data_analysis}.
The dataset consists of $1{,}103{,}602$ unique users, $125{,}541$ distinct news articles, and a total of $37{,}966{,}985$ impression logs. 

The length distributions of the articles' titles, abstracts, and bodies are shown in Figures \ref{fig:title_len}-\ref{fig:body_len}. Titles are generally concise, with an average length of just $6.6$ words. In contrast, abstracts and bodies are significantly longer with $17.3$ and $362.2$ words, respectively. 
The peaks near zero in Figures \ref{fig:subtitle_len}-\ref{fig:body_len} are due to the fact that some articles may only contain a \textit{title}.
\Cref{fig:inview_len} displays the distribution of in view articles -- those deemed visible to the user during an impression. As noted, the distribution is heavily skewed toward a small number of viewed articles per impression, a characteristic inherent to Ekstra Bladet's layout that presents its own set of challenges. In addition, as described in Section \ref{sec:dataset-construction}, active users are defined as those who had at least 5 articles in view; hence, there are no impressions with fewer present in the dataset.
In \cref{fig:category_distribution}, the distribution of article categories is shown. The categories \textit{entertainment} and \textit{news} together account for $45\%$ of the catalog. When \textit{crime} and \textit{sport} are included, these four categories collectively represent $78\%$ of all articles.

\begin{table}[tbp]
    \centering
    \caption{
    Detailed statistics of EB-NeRD.
    }
    \label{tab:EB-dataset-overview-table}    
    \begin{tabular}{  l   l  }
        \toprule
        \# News & $125{,}541$\\
        \# Users &  $1{,}103{,}602$\\
        \# Impressions  & $37{,}966{,}985$\\
        \# News categories & $32$\\   
        \# News subcategories & $263$\\   
        Avg.\@ NP-ratio & $10.5$\\
        Avg.\@ impression per user & $34.4$\\
        Avg.\@ title len. (words) & $6.6\pm2.5$\\
        Avg.\@ abstract len. (words) & $17.3\pm8.6$\\
        Avg.\@ body len. (words) & $363.2\pm306.2$\\
        \bottomrule
    \end{tabular}
\end{table}

\section{Benchmarking and Analysis}
EB-NeRD served as the benchmark dataset for the ACM Conference on Recommender Systems (RecSys) annual Challenge \citep{kruse2024recsys}\footnote{\url{https://recsys.acm.org/recsys24/challenge}}. 
The RecSys '24 Challenge focused on news recommendation, aiming to address both the technical and normative challenges involved in designing effective and responsible recommender systems for news publishing. 
This paper, like the challenge, will apply EB-NeRD to two use cases: \textit{ranking} and \textit{beyond-accuracy}. 
While, \citet{kruse2024recsys} provided an in-depth analysis of the ranking task, the beyond-accuracy aspect received only brief attention. Therefore, this paper will take the opposite approach, offering a concise analysis of the ranking task while exploring the beyond-accuracy aspects in greater detail.

\subsection{Methods}
\label{sub:methods}
Most solutions in the RecSys '24 Challenge leveraged ensemble methods, combining various models such as gradient boosting decision trees (GBDT) \citep{friedman2001greedy} -- using frameworks like LightGBM \citep{ke2017lightgbm}, CatBoost \citep{catboost_2018}, and XGBoost \citep{XGBoost_2016} -- and deep learning models, including Transformer-based models \citep{transformer_2017}. 
Hence, whereas the RecSys '24 challenge focused on evaluating the whole solution (e.g., the ensemble), we will isolate the individual ensemble components and evaluate the performance of the individual recommender systems and how they may influence a news flow.
We will briefly introduce the teams and their methods. Although teams may have used the same models or frameworks, their data pipelines and feature engineering approaches varied.
\begin{itemize}
    \item 
        \textit{:D} \citep{Fujikawa2024}, the winning team, employed a transformer-based model (\textbf{Transformer}) alongside two implementations of GBDT: \textbf{LightGBM} and \textbf{CatBoost}.
        To distinguish the methods from team \textit{:D}, we denote them as \textbf{LightGBM}$_{\mathbf{:D}}$ and \textbf{CatBoost}$_{\mathbf{:D}}$.
        Additionally, they introduced time-aware feature engineering methods and data-splitting strategies to address the temporal nature of news articles and enhance the model's generalization.
    \item 
        \textit{BlackPearl} \citep{Xue2024}, the runner-up team, implemented two GBDT models using CatBoost: pairwise ranking loss (\textbf{CatBoost}$_{\mathbf{PL}}$) and query-level loss (\textbf{CatBoost}$_{\mathbf{QL}}$). Additionally, they developed a novel neural network called Hierarchical User Interest Modeling (\textbf{HUIM}), which captures both long-term, stable user interests and short-term, rapidly changing preferences.   
    \item      
        \textit{FuxiCTR}\footnote{\url{https://github.com/reczoo/RecSys2024_CTR_Challenge}} \citep{FuxiCTR_21, FuxiCTR_22} implemented the Deep Interest Network (\textbf{DIN}) model \citep{din_2018}, which dynamically models user interests based on their historical behavior. The model employs an attention mechanism to highlight relevant interactions with the current item. The core problem DIN aims to address is ad recommendation.
    \item
        \textit{Intel\_recsys}\footnote{\url{https://github.com/yflyl613/Tiny-NewsRec}} implemented the Neural News Recommendation with Multi-Head Self-Attention (\textbf{NRMS}) model \citep{wu2019-nrms}. 
        NRMS leverages multi-head self-attention to learn user representations from their click history and to capture the relationships between news articles. 
        The core problem NRMS aims to address is news recommendation.
\end{itemize}

\subsection{Ranking}
The primary objective of the RecSys '24 Challenge followed a traditional recommendation setup \citep{Wu2020MIND}, where participants were tasked with ranking a set of articles from most to least likely for a user to click on, using various data points such as the user's click history, session details, user metadata, and the content of the articles for each impression. The ranked articles were then compared to the actual user clicks.
To evaluate the effectiveness of the recommendations in terms of ranking standard metrics were used, including Area Under the Curve (AUC), Mean Reciprocal Rank (MRR), and normalized Discounted Cumulative Gain (nDCG) up to position $K$, denoted $@K$ \citep{Wu2020MIND, Wu2023}. 

\subsubsection{Analysis}
\Cref{tab:ranking_metrics} presents the results for the ranking task. A notable observation is the presence of popularity bias: both the \textit{Clicks} ($59.70$) and \textit{Read-time} ($59.40$) baselines significantly outperform the \textit{Random} baseline ($49.98$). These baselines provide strong benchmarks for model performance. For instance, the NRMS model only slightly surpasses the strategy of predicting the most popular articles. Additionally, although both teams \textit{:D} and \textit{BlackPearl} implemented GBDT using CatBoost, their performances differ. This variation is not surprising, as each team has its own data pipeline and approach to feature engineering. 
\begin{table}[btp]
    \centering
    \caption{
    Model evaluation using AUC, MRR, nDCG@5, and nDCG@10.
    Here, 
    (1) methods from the winning team \textit{:D},
    (2) methods from the runner-up team \textit{BlackPearl}, 
    (3) the DIN model from the team \textit{FuxiCTR}, 
    (4) the NRMS model from the team \textit{Intel\_recsys}, and 
    (5) four baselines: 
        articles with the most clicks, 
        articles with the highest read-time consumption, 
        articles with the highest in view rate, and 
        a random selection.
    }
    \label{tab:ranking_metrics}
    \begin{tabular}{cl|cccc}
    \hline
     & Method         & \multicolumn{1}{c}{AUC}         & \multicolumn{1}{c}{MRR}       & \multicolumn{1}{c}{nDCG@5}   & \multicolumn{1}{c}{nDCG@10} \\ \hline
    \multirow{3}{*}{1} 
        & Transformer                   & \textbf{88.64}& \textbf{72.28} & \textbf{78.27} & \textbf{79.10} \\ 
        & CatBoost\textsubscript{:D}    & 88.05 & 70.79 & 76.75 & 77.89 \\
        & LightGBM\textsubscript{:D}    & 88.17 & 71.04 & 77.99 & 78.09 \\ \hline \multirow{3}{*}{2}
        & HUIM                          & 87.29 & 70.05 & 76.21 & 77.35 \\ 
        & CatBoost\textsubscript{QL}    & 87.89 & 71.12 & 77.13 & 78.18 \\ 
        & CatBoost\textsubscript{PL}    & 87.54 & 70.33 & 76.48 & 77.58 \\ \hline 
    3   & DIN                           & 71.53	& 49.33 & 55.08 & 59.58 \\ \hline 
    4   & NRMS                          & 61.03 & 39.75	& 44.45 & 51.24 \\ \hline \multirow{4}{*}{5}
        & Clicks                        & 59.70 & 37.74 & 42.36 & 49.65 \\
        & Read-time                     & 59.40 & 37.10 & 41.79 & 49.13 \\
        & In view                       & 54.50 & 32.39 & 36.48 & 44.74 \\
        & Random                        & 49.98 & 31.56 & 34.89 & 43.38 \\
    
    \hline
    \end{tabular}
\end{table}

\subsection{Beyond-Accuracy}
When implemented at scale, news recommender systems will play a decisive role in organizing the news that readers are exposed to and ultimately consume. This represents a deeply editorial function, with the potential to alter the editorial profile of a news brand and influence its democratic functions \citep{helberger2019}. For this reason, most mission-driven news publishers are highly invested in understanding the impact of these systems on their news flow and finding ways to guide personalized news recommendations to align with their editorial values.
EB-NeRD provides rich content metadata (e.g., news category, topics, entities, sentiment) along with key demographic attributes for some users (gender, age, location) -- see Section \cref{sec:dataset-analysis} -- enabling the evaluation of how different recommendation methods affect both overall news flow and specific reader groups. This allows for the development of models that optimize beyond simple accuracy in an offline setting.

To encourage exploration of ways to evaluate and optimize beyond-accuracy objectives using EB-NeRD, the RecSys '24 Challenge took an initial step by evaluating all submissions with an extensive, albeit preliminary, framework for beyond-accuracy analysis \citep{kruse2024recsys}.
This framework included four traditional beyond-accuracy metrics well-known in the information retrieval literature \citep{Smyth2001_intralistdiversity, ge2010_serendipity_coverage, Kaminskas2016} and re-interpreted them within the context of news publishing: intralist-diversity, novelty, serendipity, and coverage. Additionally, descriptive statistics on the distribution of recommended articles by category, topic, and sentiment were provided.

\subsubsection{Dataset}
For the evaluation, we generated a subset of $200{,}000$ users such that $50{,}000$ of them have an account. Selecting users with an account allowed for the segmentation of specific reader groups by gender, age, and geography. Among the sampled users with an account, $33{,}434$ disclosed their gender, with $28{,}404$ identifying as men and $5{,}030$ as women. 
Each user ID had a single impression specifically designed for beyond-accuracy analysis, meaning that all impressions contained the same static set of in view articles -- namely, the $250$ newly published articles available during the test period, which were unseen by any user.

\subsubsection{Analysis}
The preliminary beyond-accuracy evaluation revealed significant variation in how different recommendation methods influenced the news flow, including differences in the news flow for men and women. 
\Cref{tab:sentiment_distribution_classic_ba} presents the the classic beyond-accuracy metrics along with distribution of article sentiment, while \cref{tab:category_distribution} shows the distribution of recommended articles by category for selected submissions to the RecSys '24 Challenge.
The tables include both a selection of recommendation methods (see \cref{sub:methods}) and the aggregated results from the top 32 submissions ($\text{AUC} > 60$), top 48 submissions ($\text{AUC} > 52.5$), and the best submission from each of the 78 teams, all assessed by AUC.

These results highlights that different recommendation methods have markedly different impacts on news flow. For instance, NRMS recommended a low proportion of articles in the \textit{news} category ($10.2\%$) but a very high proportion of \textit{auto-generated} content ($44.6\%$). In contrast, CatBoost\textsubscript{:D} recommended a large proportion of \textit{news} articles ($45.7\%$) and almost no \textit{auto-generated} content ($0.02\%$).
Sentiment also varied significantly, with CatBoost\textsubscript{QL} recommending the highest share of \textit{negatively-sentimented} articles ($55.9\%$) while NRMS had a much lower share ($30.6\%$).

Additionally, several methods produced different news flows for men and women. Across the majority of RecSys Challenge submissions, \textit{sports} articles were underrepresented in the news flow for women by $6.3$ percentage points in the top 32, $4.7$ percentage points in the top 48, and $3.1$ percentage points in the top 78. 
Conversely, \textit{entertainment} articles were overrepresented by 2.6, 1.8, and 1.2 percentage points, respectively.
There was also a slight tendency for articles with negative sentiment to be overrepresented in the news flow for women by $2.6$ percentage points (top 32), $1.8$ percentage points (top 48), and $1.1$ percentage points (top 78). Certain methods exhibited even larger gender differences, such as the Transformer model, which recommended significantly more \textit{crime} content ($7.1$ percentage points) and less sports content ($11.3$ percentage points) for women compared to men.

There were also substantial differences in the performance of different recommendation methods on traditional beyond-accuracy metrics. 
For instance, diversity scores ranged from $0.791$ (Cat\-Boost\textsubscript{:D}) to $0.588$ (HUIM), serendipity varied from $0.786$ (NRMS) to $0.730$ (DIN), coverage spanned from a low of $0.152$ (CatBoost\textsubscript{QL}) to a high of $0.884$ (NRMS), and novelty ranged from $11.956$ (NRMS) to $3.368$ (CatBoost\textsubscript{:D}). However, these differences did not appear to extend to demographic categories such as gender.

\subsubsection{Discussion}
While this beyond-accuracy evaluation clearly demonstrates that different recommendation methods produce distinct news flows and that many methods create different flows for men and women, the preliminary evaluation framework we use has some important limitations.

First, it includes only a limited set of beyond-accuracy objectives derived from information retrieval literature, which may not fully reflect the most relevant beyond-accuracy goals in the news publishing domain.
A more comprehensive evaluation would require the development of new metrics that align more closely with the democratic functions of news publishers and their editorial missions.
Second, the socially relevant effects of news recommendation can only be fully assessed in real-world, online settings over extended periods. For instance, evaluating the diversity of news users are exposed to, or the level of fragmentation caused by personalization, requires multiple interactions over time. The static nature of our test's candidate list cannot capture these dynamic effects.
Third, each method was only run once, and as mentioned in \cref{sub:methods}, the underlying data pipelines and feature engineering approaches differed between teams. For example, while both \textit{:D} and \textit{BlackPearl} used the CatBoost implementation of GBDT, their models were trained differently. 
Experiments should be run multiple times using shared data pipelines, with only the seed being changed, to observe variations in performance and ensure comparability.
Finally, aligning news recommendation systems with the normative goals of news publishers ultimately requires formulating specific beyond-accuracy objectives that these systems can optimize for \citep{Heitz2024}. 
Such goals were not included in the RecSys '24 challenge, as more research is needed on beyond-accuracy metrics and their correlation with socially relevant outcomes to formulate meaningful targets to optimize for. 

Our hope is that by providing the EB-NeRD dataset, which supports extensive beyond-accuracy evaluation, and offering a preliminary framework for analysis, we inspire further research into the normative alignment of news recommendation systems that will ultimately enable us to integrate editorially and socially relevant goals in the design of news recommenders. 

\begin{table*}[tbp]
    \centering
    \caption{
    Beyond-accuracy metrics (diversity, serendipity, coverage, and novelty) and sentiment distribution in recommended articles for all users, expressed as a percentage [\%].
    The difference between the distribution for men and women is shown in percentage points in parentheses ``( $\cdot$ )'', where ``$-$'' indicates more recommendations for men, and ``$+$'' indicates more for women.
    Here, 
        (1) methods from the winning team \textit{:D}, 
        (2) methods from the runner-up team \textit{BlackPearl}, 
        (3) the DIN model from the team \textit{FuxiCTR}, 
        (4) the NRMS model from the team \textit{Intel\_recsys}, 
        (5) the random baseline, and 
        (6) the mean aggregated results from the top 32, 48, and 72, where $\pm$ represents the standard deviation.
    }
    \label{tab:sentiment_distribution_classic_ba}
    \begin{tabular}{cl|c|c|c|c|ccc}
    \hline
     & & Diversity & Serendipity & Coverage & Novelty & Negative & Neutral & Positive \\
    \hline
    \multirow{3}{*}{1}
        & Transformer                   & 0.70 (+0.01)    & 0.76 (-0.01)  & 0.56 (-0.05)  & 4.3 (+0.1)   & 41.8 (+5.1) & 32.2 (+1.0)  & 26.0 (-6.1)   \\ 
        & CatBoost\textsubscript{:D}    & 0.79 (+0.00)    & 0.77 (+0.00)  & 0.48 (-0.05)  & 3.4 (-0.0)   & 50.9 (+4.6) & 26.3 (-2.9)  & 22.8 (-1.7)   \\ 
        & LightGBM\textsubscript{:D}    & 0.71 (+0.03)    & 0.76 (+0.01)  & 0.67 (-0.04)  & 5.6 (+0.0)   & 46.4 (+2.9) & 29.8 (+4.6)  & 23.8  (-7.5)  \\ \hline\multirow{3}{*}{2}
        & HUIM                          & 0.59 (+0.03)    & 0.78 (-0.00)  & 0.52 (-0.07)  & 10.1 (-0.3)  & 30.8 (+5.5) & 28.8 (-1.0)  & 40.4 (-4.5)   \\
        & CatBoost\textsubscript{QL}    & 0.71 (+0.00)    & 0.76 (-0.01)  & 0.15 (-0.01)  & 4.7 (-0.1)   & 55.9 (-2.6) & 37.7 (+4.3)  & 6.4 (-1.8)    \\
        & CatBoost\textsubscript{PL}    & 0.63 (+0.00)    & 0.76 (-0.01)  & 0.34 (-0.03)  & 9.6 (-0.8)   & 43.7 (+3.2) & 15.5 (+4.9)  & 40.8 (-8.1)   \\\hline
    3   & DIN                           & 0.60 (+0.00)    & 0.73 (+0.00)  & 0.62 (-0.02)  & 5.5 (-0.3)   & 45.6 (+6.6) & 29.3 (-4.2)  & 25.1 (-2.4)   \\ \hline
    4   & NRMS                          & 0.61 (-0.01)    & 0.79 (+0.00)  & 0.88 (-0.14)  & 12.0 (+0.3)  & 30.6 (+2.0) & 39.1 (-0.7)  & 30.3 (-1.3)   \\ \hline
    5   & Random                        & 0.76 (+0.00)    & 0.81 (+0.00)  & 1.00 (+0.00)  & 11.1 (-0.0)  & 39.6 (-0.4) & 29.2 (-0.1)  & 31.2 (+0.5)   \\ \hline
    \multirow{3}{*}{6}
        & $\text{Top}@32_\mu$   & $\underset{\pm 0.08}{0.72}$ (-0.00) & $\underset{\pm 0.03}{0.79}$ (-0.00) & $\underset{\pm 0.29}{0.35}$ (-0.03) & $\underset{\pm 3.2}{7.6}$ (-0.0) & $\underset{\pm 0.12}{46.6}$ $(+2.6)$ & $\underset{\pm 0.08}{26.3}$ $(+0.4)$  & $\underset{\pm 0.11}{27.1}$ $(-2.9)$ \\
        & $\text{Top}@48_\mu$   & $\underset{\pm 0.18}{0.70}$ (-0.00) & $\underset{\pm 0.16}{0.76}$ (-0.00) & $\underset{\pm 0.27}{0.26}$ (-0.02) & $\underset{\pm 4.0}{7.3}$ (+0.0) & $\underset{\pm 0.19}{42.0}$ $(+1.8)$ & $\underset{\pm 0.12}{26.2}$ $(+0.2)$  & $\underset{\pm 0.16}{27.8}$ $(-2.1)$ \\
        & $\text{Top}@78_\mu$   & $\underset{\pm 0.08}{0.74}$ (-0.00) & $\underset{\pm 0.02}{0.80}$ (-0.00) & $\underset{\pm 0.42}{0.51}$ (-0.02) & $\underset{\pm 3.4}{8.9}$ (+0.0) & $\underset{\pm 0.14}{42.7}$ $(+1.1)$ & $\underset{\pm 0.09}{27.5}$ $(+0.2)$  & $\underset{\pm 0.13}{29.8}$ $(+1.3)$ \\
    \hline
    \end{tabular}
\end{table*}

\begin{table*}[tbp]
    \centering
    \caption{
    News category distribution in recommended articles for all users, expressed as a percentage [\%].
    The categories include crime (CRM), news (NWS), sports (SPT), entertainment (ENT), personal finance (PFI), auto-generated content (AGC), and miscellaneous (MSC).
    The difference between the distribution for men and women is shown in percentage points in parentheses ``( $\cdot$ )'', where ``$-$'' indicates more recommendations for men, and ``$+$'' indicates more for women.
    Here, 
        (1) methods from the winning team \textit{:D}, 
        (2) methods from the runner-up team \textit{BlackPearl}, 
        (3) the DIN model from the team \textit{FuxiCTR}, 
        (4) the NRMS model from the team \textit{Intel\_recsys}, 
        (5) the random baseline, and 
        (6) the mean aggregated results from the top 32, 48, and 72, where $\pm$ represents the standard deviation.
    }
    \label{tab:category_distribution}
    \begin{tabular}{cl|ccccccc}
    \hline 
    & & CRM & NWS & SPT & ENT & PFI & AGC & MSC \\ 
    \hline
    \multirow{3}{*}{1}
        & Transformer                   & 13.0 (+7.1)   & 13.0  (-0.6)  & 31.0  (-14.5)  & 16.6  (+11.3)  & 12.9 (-4.1)    & 0.7   (+0.2)   & 12.8 (+0.6)   \\ 
        & CatBoost\textsubscript{:D}    & 17.8 (+1.3)   & 45.7  (+4.4)  & 13.1  (-6.5)   & 2.5   (+2.0)   & 19.4 (-0.7)    & 0.0   (+0.0)   & 1.5  (-0.5)   \\
        & LightGBM\textsubscript{:D}    & 7.7  (+3.7)   & 40.3  (+5.0)  & 26.1  (-13.4)  & 7.2   (+4.1)   & 10.7 (-1.1)    & 4.3   (+0.6)   & 3.7  (+1.2)   \\ \hline \multirow{3}{*}{2}
        & HUIM                          & 1.9  (+1.8)   & 0.6   (-0.3)  & 31.9  (+1.2)   & 16.3  (+0.4)   & 2.1  (-1.3)    & 33.0  (-2.0)   & 13.1 (-0.3)   \\
        & CatBoost\textsubscript{QL}    & 4.0  (-1.6)   & 9.1   (-1.6)  & 2.0   (-1.3)   & 30.1  (+6.4)   & 14.4 (-1.5)    & 1.9   (-0.7)   & 38.5 (+0.2)   \\
        & CatBoost\textsubscript{PL}    & 1.5  (+0.4)   & 15.2  (-0.3)  & 1.4   (-0.5)   & 8.5   (+6.7)   & 12.7 (-2.1)    & 36.6  (-5.6)   & 24.1 (+1.4)   \\ \hline
    3   & DIN                           & 5.8  (+2.0)   & 22.6  (-0.8)  & 38.0  (-14.3)  & 12.7  (+7.0)   & 4.2  (+0.4)    & 0.6   (+0.2)   & 16.1 (+5.3)   \\ \hline
    4   & NRMS                          & 5.4  (-0.2)   & 10.2  (+2.7)  & 26.5  (-10.2)  & 2.2   (+1.7)   & 4.3  (-0.1)    & 44.6  (+2.2)   & 6.8  (+3.9)   \\ \hline
    5   & Random                        & 9.6  (-0.1)   & 16.0  (-0.1)  & 16.8  (-0.1)   & 8.4   (+0.0)   & 3.6  (+0.2)    & 38.0  (+0.1)   & 7.2  (-0.0)   \\ \hline
    \multirow{3}{*}{6}
        & $\text{Top}@32_\mu$ & $\underset{\pm 6.5}{9.2}$ $(+1.9)$  & $\underset{\pm 13.9}{16.8}$ $(+0.9)$  & $\underset{\pm 13.0}{25.2}$ $(-6.3)$  & $\underset{\pm 11.7}{15.9}$ $(+2.6)$  & $\underset{\pm 8.3}{5.5}$ $(-0.4)$  & $\underset{\pm 17.8}{14.0}$ $(+0.4)$  & $\underset{\pm 0.0}{11.4}$ $(+0.6)$ \\
        & $\text{Top}@48_\mu$ & $\underset{\pm 7.5}{9.4}$ $(+1.2)$  & $\underset{\pm 14.5}{15.3}$ $(+0.8)$  & $\underset{\pm 15.5}{25.4}$ $(-4.7)$  & $\underset{\pm 13.1}{13.8}$ $(+1.8)$  & $\underset{\pm 7.8}{5.5}$ $(-0.2)$  & $\underset{\pm 21.4}{14.8}$ $(+0.5)$  & $\underset{\pm 0.0}{10.6}$ $(+0.4)$ \\
        & $\text{Top}@72_\mu$ & $\underset{\pm 6.3}{9.6}$ $(+0.7)$  & $\underset{\pm 13.6}{17.4}$ $(+0.5)$  & $\underset{\pm 13.1}{22.3}$ $(-3.1)$  & $\underset{\pm 10.8}{12.2}$ $(+1.2)$  & $\underset{\pm 6.3}{4.7}$ $(-0.1)$  & $\underset{\pm 20.4}{23.5}$ $(+0.4)$  & $\underset{\pm 0.0}{9.3}$ $(+0.4)$ \\
    \hline 
    \end{tabular}
\end{table*}

\section{Conclusion}
In this paper, we present EB-NeRD, a large-scale dataset for news recommendation, constructed from the user behavior logs of Ekstra Bladet. 
The dataset comprises over $1$ million users and more than $125{,}000$ Danish news articles, each enriched with detailed textual content such as titles, abstracts, and full bodies.
Additionally, through the RecSys '24 Challenge, we facilitated and conducted extensive experiments and analysis on this dataset. The results underscore the critical importance of accurately understanding news content and modeling user interests for effective news recommendation. 
While our initial analysis of beyond-accuracy metrics is limited, it suggests that different recommender systems can have significantly varying impacts on the news flow. We hope these findings will encourage further research on both the technical and normative challenges associated with recommender systems in news publishing.

\section{Future Directions}
EB-NeRD offers a broad range of research opportunities, including the development of advanced methods for news and user modeling, enhancing the diversity, fairness, and explainability of news recommendations, and exploring privacy-preserving techniques in news recommendation. 
We have open-sourced this comprehensive dataset to facilitate and encourage research on both the technical and normative challenges associated with recommender systems in news publishing. 
The dataset is intended to serve as a foundation for advanced research in these areas. 
While we anticipate innovative applications of the dataset, we propose that at least three lines of inquiry will be particularly valuable and contribute significantly to ongoing research efforts.
\begin{itemize}
    \item \textbf{Development of scalable recommender systems for real-world applications}: 
    In evaluating the benchmark methods in this paper, it is important to acknowledge a significant limitation: the predominant reliance on complex ensemble techniques.
    While these approaches are effective for achieving high scores and winning machine learning competitions \citep{Jannach2020_recsys_winning_solutions}, they often involve ensembles of models that exploit specific characteristics of static datasets \citep{kruse2024recsys}. 
    However, such solutions may not be feasible for real-world deployment due to their computational demands and difficulty in adapting to real-time environments. This highlights the need for developing more practical and scalable solutions that balance performance with operational efficiency.

    \item \textbf{Development of more advanced recommender systems that specifically tackle the core challenges of news recommendation}: 
    As argued by \citet{Wu2023}, one of the most promising ways to improve news recommendations is by developing advanced recommender systems that address the unique technical challenges of the news domain. 
    Although we only had one model specifically designed to address the core problem of news recommendation, which was NRMS, EB-NeRD presents opportunities to develop advanced recommender systems that tackle these fundamental challenges.
    Although we only evaluated two models specifically designed for news recommendation, i.e., NRMS and HUIM, EB-NeRD presents opportunities to develop systems that tackle these core challenges. 
    Additionally, it remains to be seen whether the performance achieved by \citet{Wu2020MIND} can be generalized to classical digital news publishers, like those in EB-NeRD and Adressa, which have news consumption and content profiles that differ significantly from those of news aggregators, such as MIND.

    \item \textbf{Development of more advanced strategies for value alignment of recommender systems}: 
    Development of more advanced strategies for value alignment of recommender systems: As argued above \citep[see also, e.g.,][]{helberger2019, lu2020_editoral_values, NORMalize_23_workshop}, there is a need for alternative evaluation methods for news recommendations that examine value alignment with editorial values as well as different models of news publishers role in democracy. While we only provide evaluation against traditional, generic beyond-accuracy objectives in this paper, EB-NeRD allows for the development of alternative beyond-accuracy metrics and strategies for optimization against them that are more closely aligned with domain specific journalistic or democratic values. Such research is already ongoing \citep[see, e.g.,][]{Vrijenhoek2021_radio}, and we believe that use of the dataset provided for this purpose will provide a fruitful avenue for research with significant industrial relevance.
    Such insights are essential for publishers, and we hope they will inspire further work on beyond-accuracy objectives to support value alignment, which is an emerging research agenda \citep[see, e.g.,][]{helberger2019, lu2020_editoral_values, Vrijenhoek2021_radio, Stray2023}.
\end{itemize}

\begin{acks}
We would like to extend our gratitude to the challenge participants and the RecSys organizers for their engagement and support. We also wish to acknowledge our employers and funding bodies, including 
Ekstra Bladet, 
JP/Politikens Media Group, 
Technical University of Denmark, 
Copenhagen Business School, 
Innovation Foundation Denmark (grant number 1044-00058B), 
and the Platform Intelligence in News-Project (grant number 0175-00014B). 
The work of Marco Polignano is supported by the PNRR project FAIR - Future AI Research (PE00000013), Spoke 6 - Symbiotic AI (CUP H97G22000210007) under the NRRP MUR program funded by the NextGenerationEU.
\end{acks}

\bibliographystyle{ACM-Reference-Format}
\bibliography{main}


\begin{thebibliography}{52}


\ifx \showCODEN    \undefined \def \showCODEN     #1{\unskip}     \fi
\ifx \showDOI      \undefined \def \showDOI       #1{#1}\fi
\ifx \showISBNx    \undefined \def \showISBNx     #1{\unskip}     \fi
\ifx \showISBNxiii \undefined \def \showISBNxiii  #1{\unskip}     \fi
\ifx \showISSN     \undefined \def \showISSN      #1{\unskip}     \fi
\ifx \showLCCN     \undefined \def \showLCCN      #1{\unskip}     \fi
\ifx \shownote     \undefined \def \shownote      #1{#1}          \fi
\ifx \showarticletitle \undefined \def \showarticletitle #1{#1}   \fi
\ifx \showURL      \undefined \def \showURL       {\relax}        \fi
\providecommand\bibfield[2]{#2}
\providecommand\bibinfo[2]{#2}
\providecommand\natexlab[1]{#1}
\providecommand\showeprint[2][]{arXiv:#2}

\bibitem[Abel et~al\mbox{.}(2011)]%
        {Abel2011}
\bibfield{author}{\bibinfo{person}{Fabian Abel}, \bibinfo{person}{Qi Gao}, \bibinfo{person}{Geert~Jan Houben}, {and} \bibinfo{person}{Ke Tao}.} \bibinfo{year}{2011}\natexlab{}.
\newblock \showarticletitle{{Analyzing user modeling on Twitter for personalized news recommendations}}.
\newblock \bibinfo{journal}{\emph{Lecture Notes in Computer Science (including subseries Lecture Notes in Artificial Intelligence and Lecture Notes in Bioinformatics)}}  \bibinfo{volume}{6787 LNCS} (\bibinfo{year}{2011}), \bibinfo{pages}{1--12}.
\newblock
\showISBNx{9783642223617}
\showISSN{03029743}
\urldef\tempurl%
\url{https://doi.org/10.1007/978-3-642-22362-4_1}
\showDOI{\tempurl}


\bibitem[Acharya et~al\mbox{.}(2023)]%
        {acharya2023}
\bibfield{author}{\bibinfo{person}{Arkadeep Acharya}, \bibinfo{person}{Brijraj Singh}, {and} \bibinfo{person}{Naoyuki Onoe}.} \bibinfo{year}{2023}\natexlab{}.
\newblock \showarticletitle{LLM Based Generation of Item-Description for Recommendation System}. In \bibinfo{booktitle}{\emph{Proceedings of the 17th ACM Conference on Recommender Systems}} (Singapore, Singapore) \emph{(\bibinfo{series}{RecSys '23})}. \bibinfo{publisher}{Association for Computing Machinery}, \bibinfo{address}{New York, NY, USA}, \bibinfo{pages}{1204–1207}.
\newblock
\showISBNx{9798400702419}
\urldef\tempurl%
\url{https://doi.org/10.1145/3604915.3610647}
\showDOI{\tempurl}


\bibitem[An et~al\mbox{.}(2019)]%
        {an2019-lstur}
\bibfield{author}{\bibinfo{person}{Mingxiao An}, \bibinfo{person}{Fangzhao Wu}, \bibinfo{person}{Chuhan Wu}, \bibinfo{person}{Kun Zhang}, \bibinfo{person}{Zheng Liu}, {and} \bibinfo{person}{Xing Xie}.} \bibinfo{year}{2019}\natexlab{}.
\newblock \showarticletitle{Neural News Recommendation with Long- and Short-term User Representations}. In \bibinfo{booktitle}{\emph{Proceedings of the 57th Annual Meeting of the Association for Computational Linguistics}}. \bibinfo{publisher}{Association for Computational Linguistics}, \bibinfo{address}{Florence, Italy}, \bibinfo{pages}{336--345}.
\newblock
\urldef\tempurl%
\url{https://doi.org/10.18653/v1/P19-1033}
\showDOI{\tempurl}


\bibitem[Chen and Guestrin(2016)]%
        {XGBoost_2016}
\bibfield{author}{\bibinfo{person}{Tianqi Chen} {and} \bibinfo{person}{Carlos Guestrin}.} \bibinfo{year}{2016}\natexlab{}.
\newblock \showarticletitle{XGBoost: A Scalable Tree Boosting System}. In \bibinfo{booktitle}{\emph{Proceedings of the 22nd ACM SIGKDD International Conference on Knowledge Discovery and Data Mining}} (San Francisco, California, USA) \emph{(\bibinfo{series}{KDD '16})}. \bibinfo{publisher}{Association for Computing Machinery}, \bibinfo{address}{New York, NY, USA}, \bibinfo{pages}{785–794}.
\newblock
\showISBNx{9781450342322}
\urldef\tempurl%
\url{https://doi.org/10.1145/2939672.2939785}
\showDOI{\tempurl}


\bibitem[Das et~al\mbox{.}(2007)]%
        {Das2007}
\bibfield{author}{\bibinfo{person}{Abhinandan~S. Das}, \bibinfo{person}{Mayur Datar}, \bibinfo{person}{Ashutosh Garg}, {and} \bibinfo{person}{Shyam Rajaram}.} \bibinfo{year}{2007}\natexlab{}.
\newblock \showarticletitle{Google News Personalization: Scalable Online Collaborative Filtering}. In \bibinfo{booktitle}{\emph{Proceedings of the 16th International Conference on World Wide Web}} (Banff, Alberta, Canada) \emph{(\bibinfo{series}{WWW '07})}. \bibinfo{publisher}{Association for Computing Machinery}, \bibinfo{address}{New York, NY, USA}, \bibinfo{pages}{271–280}.
\newblock
\showISBNx{9781595936547}
\urldef\tempurl%
\url{https://doi.org/10.1145/1242572.1242610}
\showDOI{\tempurl}


\bibitem[Feng et~al\mbox{.}(2020)]%
        {Feng2020}
\bibfield{author}{\bibinfo{person}{Chong Feng}, \bibinfo{person}{Muzammil Khan}, \bibinfo{person}{Arif~Ur Rahman}, {and} \bibinfo{person}{Arshad Ahmad}.} \bibinfo{year}{2020}\natexlab{}.
\newblock \showarticletitle{{News Recommendation Systems-Accomplishments, Challenges Future Directions}}.
\newblock \bibinfo{journal}{\emph{IEEE Access}}  \bibinfo{volume}{8} (\bibinfo{year}{2020}), \bibinfo{pages}{16702--16725}.
\newblock
\showISSN{21693536}
\urldef\tempurl%
\url{https://doi.org/10.1109/ACCESS.2020.2967792}
\showDOI{\tempurl}


\bibitem[Friedman(2001)]%
        {friedman2001greedy}
\bibfield{author}{\bibinfo{person}{Jerome~H Friedman}.} \bibinfo{year}{2001}\natexlab{}.
\newblock \showarticletitle{Greedy function approximation: a gradient boosting machine}.
\newblock \bibinfo{journal}{\emph{Annals of statistics}} \bibinfo{volume}{29}, \bibinfo{number}{5} (\bibinfo{year}{2001}), \bibinfo{pages}{1189--1232}.
\newblock


\bibitem[Fujikawa et~al\mbox{.}(2024)]%
        {Fujikawa2024}
\bibfield{author}{\bibinfo{person}{Kazuki Fujikawa}, \bibinfo{person}{Naoki Murakami}, {and} \bibinfo{person}{Yuki Sugawara}.} \bibinfo{year}{2024}\natexlab{}.
\newblock \showarticletitle{Enhancing News Recommendation with Transformers and Ensemble Learning}. In \bibinfo{booktitle}{\emph{Proceedings of the Recommender Systems Challenge 2024}} (Bari, Italy) \emph{(\bibinfo{series}{RecSysChallenge '24})}. \bibinfo{publisher}{Association for Computing Machinery}, \bibinfo{address}{New York, NY, USA}.
\newblock
\showISBNx{979-8-4007-1127-5/24/10}
\urldef\tempurl%
\url{https://doi.org/10.1145/3687151.3687160}
\showDOI{\tempurl}


\bibitem[Ge et~al\mbox{.}(2010)]%
        {ge2010_serendipity_coverage}
\bibfield{author}{\bibinfo{person}{Mouzhi Ge}, \bibinfo{person}{Carla Delgado-Battenfeld}, {and} \bibinfo{person}{Dietmar Jannach}.} \bibinfo{year}{2010}\natexlab{}.
\newblock \showarticletitle{Beyond Accuracy: Evaluating Recommender Systems by Coverage and Serendipity}. In \bibinfo{booktitle}{\emph{Proceedings of the Fourth ACM Conference on Recommender Systems}} (Barcelona, Spain) \emph{(\bibinfo{series}{RecSys '10})}. \bibinfo{publisher}{Association for Computing Machinery}, \bibinfo{address}{New York, NY, USA}, \bibinfo{pages}{257–260}.
\newblock
\showISBNx{9781605589060}
\urldef\tempurl%
\url{https://doi.org/10.1145/1864708.1864761}
\showDOI{\tempurl}


\bibitem[Gu et~al\mbox{.}(2016)]%
        {Gu2016}
\bibfield{author}{\bibinfo{person}{Wanrong Gu}, \bibinfo{person}{Shoubin Dong}, {and} \bibinfo{person}{Mingquan Chen}.} \bibinfo{year}{2016}\natexlab{}.
\newblock \showarticletitle{{Personalized news recommendation based on articles chain building}}.
\newblock \bibinfo{journal}{\emph{Neural Computing and Applications}} \bibinfo{volume}{27}, \bibinfo{number}{5} (\bibinfo{year}{2016}), \bibinfo{pages}{1263--1272}.
\newblock
\showISSN{09410643}
\urldef\tempurl%
\url{https://doi.org/10.1007/s00521-015-1932-x}
\showDOI{\tempurl}


\bibitem[Gulla et~al\mbox{.}(2017)]%
        {Gulla2017_adressa}
\bibfield{author}{\bibinfo{person}{Jon~Atle Gulla}, \bibinfo{person}{Lemei Zhang}, \bibinfo{person}{Peng Liu}, \bibinfo{person}{\"{O}zlem \"{O}zg\"{o}bek}, {and} \bibinfo{person}{Xiaomeng Su}.} \bibinfo{year}{2017}\natexlab{}.
\newblock \showarticletitle{The Adressa Dataset for News Recommendation}. In \bibinfo{booktitle}{\emph{Proceedings of the International Conference on Web Intelligence}} (Leipzig, Germany) \emph{(\bibinfo{series}{WI '17})}. \bibinfo{publisher}{Association for Computing Machinery}, \bibinfo{address}{New York, NY, USA}, \bibinfo{pages}{1042--1048}.
\newblock
\showISBNx{9781450349512}


\bibitem[Heitz et~al\mbox{.}(2024)]%
        {Heitz2024}
\bibfield{author}{\bibinfo{person}{Lucien Heitz}, \bibinfo{person}{Oana Inel}, {and} \bibinfo{person}{Sanne Vrijenhoek}.} \bibinfo{year}{2024}\natexlab{}.
\newblock \showarticletitle{Recommendations for the Recommenders: Reflections on Prioritizing Diversity in the RecSys Challenge}. In \bibinfo{booktitle}{\emph{Proceedings of the Recommender Systems Challenge 2024}} (Bari, Italy) \emph{(\bibinfo{series}{RecSysChallenge '24})}. \bibinfo{publisher}{Association for Computing Machinery}, \bibinfo{address}{New York, NY, USA}.
\newblock
\showISBNx{979-8-4007-1127-5/24/10}
\urldef\tempurl%
\url{https://doi.org/10.1145/3687151.3687155}
\showDOI{\tempurl}


\bibitem[Helberger(2019)]%
        {helberger2019}
\bibfield{author}{\bibinfo{person}{Natali Helberger}.} \bibinfo{year}{2019}\natexlab{}.
\newblock \showarticletitle{{On the Democratic Role of News Recommenders}}.
\newblock \bibinfo{journal}{\emph{Digital Journalism}} \bibinfo{volume}{7}, \bibinfo{number}{8} (\bibinfo{year}{2019}), \bibinfo{pages}{993--1012}.
\newblock
\urldef\tempurl%
\url{https://doi.org/10.1080/21670811.2019.1623700}
\showDOI{\tempurl}


\bibitem[Huang et~al\mbox{.}(2013)]%
        {Huang2013}
\bibfield{author}{\bibinfo{person}{Po-Sen Huang}, \bibinfo{person}{Xiaodong He}, \bibinfo{person}{Jianfeng Gao}, \bibinfo{person}{Li Deng}, \bibinfo{person}{Alex Acero}, {and} \bibinfo{person}{Larry Heck}.} \bibinfo{year}{2013}\natexlab{}.
\newblock \showarticletitle{Learning deep structured semantic models for web search using clickthrough data}. In \bibinfo{booktitle}{\emph{Proceedings of the 22nd ACM International Conference on Information \& Knowledge Management}} (San Francisco, California, USA) \emph{(\bibinfo{series}{CIKM '13})}. \bibinfo{publisher}{Association for Computing Machinery}, \bibinfo{address}{New York, NY, USA}, \bibinfo{pages}{2333–2338}.
\newblock
\showISBNx{9781450322638}
\urldef\tempurl%
\url{https://doi.org/10.1145/2505515.2505665}
\showDOI{\tempurl}


\bibitem[Ilievski and Roy(2013)]%
        {Ilija2013}
\bibfield{author}{\bibinfo{person}{Ilija Ilievski} {and} \bibinfo{person}{Sujoy Roy}.} \bibinfo{year}{2013}\natexlab{}.
\newblock \showarticletitle{Personalized News Recommendation Based on Implicit Feedback}. In \bibinfo{booktitle}{\emph{Proceedings of the 2013 International News Recommender Systems Workshop and Challenge}} (Kowloon, Hong Kong) \emph{(\bibinfo{series}{NRS '13})}. \bibinfo{publisher}{Association for Computing Machinery}, \bibinfo{address}{New York, NY, USA}, \bibinfo{pages}{10–15}.
\newblock
\showISBNx{9781450323024}
\urldef\tempurl%
\url{https://doi.org/10.1145/2516641.2516644}
\showDOI{\tempurl}


\bibitem[Jannach et~al\mbox{.}(2020)]%
        {Jannach2020_recsys_winning_solutions}
\bibfield{author}{\bibinfo{person}{Dietmar Jannach}, \bibinfo{person}{Gabriel de Souza P.~Moreira}, {and} \bibinfo{person}{Even Oldridge}.} \bibinfo{year}{2020}\natexlab{}.
\newblock \showarticletitle{Why Are Deep Learning Models Not Consistently Winning Recommender Systems Competitions Yet? A Position Paper}. In \bibinfo{booktitle}{\emph{Proceedings of the Recommender Systems Challenge 2020}} (Virtual Event, Brazil) \emph{(\bibinfo{series}{RecSysChallenge '20})}. \bibinfo{publisher}{Association for Computing Machinery}, \bibinfo{address}{New York, NY, USA}, \bibinfo{pages}{44–49}.
\newblock
\showISBNx{9781450388351}
\urldef\tempurl%
\url{https://doi.org/10.1145/3415959.3416001}
\showDOI{\tempurl}


\bibitem[Jannach and Jugovac(2019)]%
        {Jannach2019}
\bibfield{author}{\bibinfo{person}{Dietmar Jannach} {and} \bibinfo{person}{Michael Jugovac}.} \bibinfo{year}{2019}\natexlab{}.
\newblock \showarticletitle{Measuring the Business Value of Recommender Systems}.
\newblock \bibinfo{journal}{\emph{ACM Trans. Manage. Inf. Syst.}} \bibinfo{volume}{10}, \bibinfo{number}{4}, Article \bibinfo{articleno}{16} (\bibinfo{date}{dec} \bibinfo{year}{2019}), \bibinfo{numpages}{23}~pages.
\newblock
\showISSN{2158-656X}
\urldef\tempurl%
\url{https://doi.org/10.1145/3370082}
\showDOI{\tempurl}


\bibitem[Kaminskas and Bridge(2016)]%
        {Kaminskas2016}
\bibfield{author}{\bibinfo{person}{Marius Kaminskas} {and} \bibinfo{person}{Derek Bridge}.} \bibinfo{year}{2016}\natexlab{}.
\newblock \showarticletitle{Diversity, Serendipity, Novelty, and Coverage: A Survey and Empirical Analysis of Beyond-Accuracy Objectives in Recommender Systems}.
\newblock \bibinfo{journal}{\emph{ACM Trans. Interact. Intell. Syst.}} \bibinfo{volume}{7}, \bibinfo{number}{1}, Article \bibinfo{articleno}{2} (\bibinfo{date}{dec} \bibinfo{year}{2016}), \bibinfo{numpages}{42}~pages.
\newblock
\showISSN{2160-6455}
\urldef\tempurl%
\url{https://doi.org/10.1145/2926720}
\showDOI{\tempurl}


\bibitem[Karimi et~al\mbox{.}(2018)]%
        {Karimi2018}
\bibfield{author}{\bibinfo{person}{Mozhgan Karimi}, \bibinfo{person}{Dietmar Jannach}, {and} \bibinfo{person}{Michael Jugovac}.} \bibinfo{year}{2018}\natexlab{}.
\newblock \showarticletitle{{News recommender systems – Survey and roads ahead}}.
\newblock \bibinfo{journal}{\emph{Information Processing and Management}} \bibinfo{volume}{54}, \bibinfo{number}{6} (\bibinfo{year}{2018}), \bibinfo{pages}{1203--1227}.
\newblock
\showISSN{03064573}
\urldef\tempurl%
\url{https://doi.org/10.1016/j.ipm.2018.04.008}
\showDOI{\tempurl}


\bibitem[Ke et~al\mbox{.}(2017)]%
        {ke2017lightgbm}
\bibfield{author}{\bibinfo{person}{Guolin Ke}, \bibinfo{person}{Qi Meng}, \bibinfo{person}{Thomas Finley}, \bibinfo{person}{Taifeng Wang}, \bibinfo{person}{Wei Chen}, \bibinfo{person}{Weidong Ma}, \bibinfo{person}{Qiwei Ye}, {and} \bibinfo{person}{Tie-Yan Liu}.} \bibinfo{year}{2017}\natexlab{}.
\newblock \showarticletitle{Lightgbm: A highly efficient gradient boosting decision tree}.
\newblock \bibinfo{journal}{\emph{Advances in neural information processing systems}}  \bibinfo{volume}{30} (\bibinfo{year}{2017}), \bibinfo{pages}{3146--3154}.
\newblock


\bibitem[Kille et~al\mbox{.}(2013)]%
        {Kille-plista}
\bibfield{author}{\bibinfo{person}{Benjamin Kille}, \bibinfo{person}{Frank Hopfgartner}, \bibinfo{person}{Torben Brodt}, {and} \bibinfo{person}{Tobias Heintz}.} \bibinfo{year}{2013}\natexlab{}.
\newblock \showarticletitle{The Plista Dataset}. In \bibinfo{booktitle}{\emph{Proceedings of the 2013 International News Recommender Systems Workshop and Challenge}} (Kowloon, Hong Kong) \emph{(\bibinfo{series}{NRS '13})}. \bibinfo{publisher}{Association for Computing Machinery}, \bibinfo{address}{New York, NY, USA}, \bibinfo{pages}{16–23}.
\newblock
\showISBNx{9781450323024}
\urldef\tempurl%
\url{https://doi.org/10.1145/2516641.2516643}
\showDOI{\tempurl}


\bibitem[Kompan and Bielikov{\'{a}}(2010)]%
        {Kompan2010}
\bibfield{author}{\bibinfo{person}{Michal Kompan} {and} \bibinfo{person}{M{\'{a}}ria Bielikov{\'{a}}}.} \bibinfo{year}{2010}\natexlab{}.
\newblock \showarticletitle{{Content-based news recommendation}}.
\newblock \bibinfo{journal}{\emph{Lecture Notes in Business Information Processing}}  \bibinfo{volume}{61 LNBIP} (\bibinfo{year}{2010}), \bibinfo{pages}{61--72}.
\newblock
\showISBNx{3642152074}
\showISSN{18651348}
\urldef\tempurl%
\url{https://doi.org/10.1007/978-3-642-15208-5_6}
\showDOI{\tempurl}


\bibitem[Koren(2008)]%
        {Koren2008}
\bibfield{author}{\bibinfo{person}{Yehuda Koren}.} \bibinfo{year}{2008}\natexlab{}.
\newblock \showarticletitle{Factorization Meets the Neighborhood: A Multifaceted Collaborative Filtering Model}. In \bibinfo{booktitle}{\emph{Proceedings of the 14th ACM SIGKDD International Conference on Knowledge Discovery and Data Mining}} (Las Vegas, Nevada, USA) \emph{(\bibinfo{series}{KDD '08})}. \bibinfo{publisher}{Association for Computing Machinery}, \bibinfo{address}{New York, NY, USA}, \bibinfo{pages}{426–434}.
\newblock
\showISBNx{9781605581934}
\urldef\tempurl%
\url{https://doi.org/10.1145/1401890.1401944}
\showDOI{\tempurl}


\bibitem[Kruse et~al\mbox{.}(2024)]%
        {kruse2024recsys}
\bibfield{author}{\bibinfo{person}{Johannes Kruse}, \bibinfo{person}{Kasper Lindskow}, \bibinfo{person}{Saikishore Kalloori}, \bibinfo{person}{Marco Polignano}, \bibinfo{person}{Claudio Pomo}, \bibinfo{person}{Abhishek Srivastava}, \bibinfo{person}{Anshuk Uppal}, \bibinfo{person}{Michael~Riis Andersen}, {and} \bibinfo{person}{Jes Frellsen}.} \bibinfo{year}{2024}\natexlab{}.
\newblock \showarticletitle{RecSys Challenge 2024: Balancing Accuracy and Editorial Values in News Recommendations}. In \bibinfo{booktitle}{\emph{Proceedings of the 18th ACM Conference on Recommender Systems}} (Bari, Italy) \emph{(\bibinfo{series}{RecSys '24})}. \bibinfo{publisher}{Association for Computing Machinery}, \bibinfo{address}{New York, NY, USA}.
\newblock
\showISBNx{979-8-4007-0505-2/24/10}
\urldef\tempurl%
\url{https://doi.org/10.1145/3640457.3687164}
\showDOI{\tempurl}


\bibitem[Lu et~al\mbox{.}(2020)]%
        {lu2020_editoral_values}
\bibfield{author}{\bibinfo{person}{Feng Lu}, \bibinfo{person}{Anca Dumitrache}, {and} \bibinfo{person}{David Graus}.} \bibinfo{year}{2020}\natexlab{}.
\newblock \showarticletitle{Beyond Optimizing for Clicks: Incorporating Editorial Values in News Recommendation}. In \bibinfo{booktitle}{\emph{Proceedings of the 28th ACM Conference on User Modeling, Adaptation and Personalization}} (Genoa, Italy) \emph{(\bibinfo{series}{UMAP '20})}. \bibinfo{publisher}{Association for Computing Machinery}, \bibinfo{address}{New York, NY, USA}, \bibinfo{pages}{145–153}.
\newblock
\showISBNx{9781450368612}
\urldef\tempurl%
\url{https://doi.org/10.1145/3340631.3394864}
\showDOI{\tempurl}


\bibitem[Misztal{-}Radecka et~al\mbox{.}(2019)]%
        {Radecka2019}
\bibfield{author}{\bibinfo{person}{Joanna Misztal{-}Radecka}, \bibinfo{person}{Dominik Rusiecki}, \bibinfo{person}{Michal Zmuda}, {and} \bibinfo{person}{Artur Bujak}.} \bibinfo{year}{2019}\natexlab{}.
\newblock \showarticletitle{Trend-responsive User Segmentation Enabling Traceable Publishing Insights. {A} Case Study of a Real-world Large-scale News Recommendation System}.
\newblock \bibinfo{journal}{\emph{CoRR}}  \bibinfo{volume}{abs/1911.11070} (\bibinfo{year}{2019}).
\newblock
\showeprint[arXiv]{1911.11070}
\urldef\tempurl%
\url{http://arxiv.org/abs/1911.11070}
\showURL{%
\tempurl}


\bibitem[Moreira et~al\mbox{.}(2018)]%
        {Moreira2018_globo}
\bibfield{author}{\bibinfo{person}{Gabriel de Souza~Pereira Moreira}, \bibinfo{person}{Felipe Ferreira}, {and} \bibinfo{person}{Adilson~Marques da Cunha}.} \bibinfo{year}{2018}\natexlab{}.
\newblock \showarticletitle{News Session-Based Recommendations Using Deep Neural Networks}. In \bibinfo{booktitle}{\emph{Proceedings of the 3rd Workshop on Deep Learning for Recommender Systems}} (Vancouver, BC, Canada) \emph{(\bibinfo{series}{DLRS 2018})}. \bibinfo{publisher}{Association for Computing Machinery}, \bibinfo{address}{New York, NY, USA}, \bibinfo{pages}{15–23}.
\newblock
\showISBNx{9781450366175}
\urldef\tempurl%
\url{https://doi.org/10.1145/3270323.3270328}
\showDOI{\tempurl}


\bibitem[Oh et~al\mbox{.}(2014)]%
        {Oh2014}
\bibfield{author}{\bibinfo{person}{Kyo-Joong Oh}, \bibinfo{person}{Won-Jo Lee}, \bibinfo{person}{Chae-Gyun Lim}, {and} \bibinfo{person}{Ho-Jin Choi}.} \bibinfo{year}{2014}\natexlab{}.
\newblock \showarticletitle{Personalized news recommendation using classified keywords to capture user preference}. In \bibinfo{booktitle}{\emph{16th International Conference on Advanced Communication Technology}}. \bibinfo{pages}{1283--1287}.
\newblock
\urldef\tempurl%
\url{https://doi.org/10.1109/ICACT.2014.6779166}
\showDOI{\tempurl}


\bibitem[Prokhorenkova et~al\mbox{.}(2018)]%
        {catboost_2018}
\bibfield{author}{\bibinfo{person}{Liudmila Prokhorenkova}, \bibinfo{person}{Gleb Gusev}, \bibinfo{person}{Aleksandr Vorobev}, \bibinfo{person}{Anna~Veronika Dorogush}, {and} \bibinfo{person}{Andrey Gulin}.} \bibinfo{year}{2018}\natexlab{}.
\newblock \showarticletitle{CatBoost: unbiased boosting with categorical features}. In \bibinfo{booktitle}{\emph{Proceedings of the 32nd International Conference on Neural Information Processing Systems}} (Montr\'{e}al, Canada). \bibinfo{publisher}{Curran Associates Inc.}, \bibinfo{address}{Red Hook, NY, USA}, \bibinfo{pages}{6639–6649}.
\newblock


\bibitem[Qi et~al\mbox{.}(2021)]%
        {Qi2021-know-interactive}
\bibfield{author}{\bibinfo{person}{Tao Qi}, \bibinfo{person}{Fangzhao Wu}, \bibinfo{person}{Chuhan Wu}, {and} \bibinfo{person}{Yongfeng Huang}.} \bibinfo{year}{2021}\natexlab{}.
\newblock \showarticletitle{Personalized News Recommendation with Knowledge-aware Interactive Matching}. In \bibinfo{booktitle}{\emph{Proceedings of the 44th International ACM SIGIR Conference on Research and Development in Information Retrieval}} (Virtual Event, Canada) \emph{(\bibinfo{series}{SIGIR '21})}. \bibinfo{publisher}{Association for Computing Machinery}, \bibinfo{address}{New York, NY, USA}, \bibinfo{pages}{61–70}.
\newblock
\showISBNx{9781450380379}
\urldef\tempurl%
\url{https://doi.org/10.1145/3404835.3462861}
\showDOI{\tempurl}


\bibitem[Raza and Ding(2022)]%
        {raza2020}
\bibfield{author}{\bibinfo{person}{Shaina Raza} {and} \bibinfo{person}{Chen Ding}.} \bibinfo{year}{2022}\natexlab{}.
\newblock \showarticletitle{News recommender system: a review of recent progress, challenges, and opportunities}.
\newblock \bibinfo{journal}{\emph{Artificial Intelligence Review}} \bibinfo{volume}{55}, \bibinfo{number}{1} (\bibinfo{year}{2022}), \bibinfo{pages}{749--800}.
\newblock
\showISBNx{1573-7462}
\urldef\tempurl%
\url{https://doi.org/10.1007/s10462-021-10043-x}
\showDOI{\tempurl}


\bibitem[Rendle(2010)]%
        {Rendle2010}
\bibfield{author}{\bibinfo{person}{Steffen Rendle}.} \bibinfo{year}{2010}\natexlab{}.
\newblock \showarticletitle{Factorization Machines}. In \bibinfo{booktitle}{\emph{2010 IEEE International Conference on Data Mining}}. \bibinfo{pages}{995--1000}.
\newblock
\urldef\tempurl%
\url{https://doi.org/10.1109/ICDM.2010.127}
\showDOI{\tempurl}


\bibitem[Rosulek(2021)]%
        {joyofcryptography}
\bibfield{author}{\bibinfo{person}{Mike Rosulek}.} \bibinfo{year}{2021}\natexlab{}.
\newblock \bibinfo{title}{The Joy of Cryptography}.
\newblock
\newblock
\urldef\tempurl%
\url{https://joyofcryptography.com}
\showURL{%
\tempurl}


\bibitem[Smyth and McClave(2001)]%
        {Smyth2001_intralistdiversity}
\bibfield{author}{\bibinfo{person}{Barry Smyth} {and} \bibinfo{person}{Paul McClave}.} \bibinfo{year}{2001}\natexlab{}.
\newblock \showarticletitle{Similarity vs. Diversity}. In \bibinfo{booktitle}{\emph{Proceedings of the 4th International Conference on Case-Based Reasoning: Case-Based Reasoning Research and Development}} \emph{(\bibinfo{series}{ICCBR '01})}. \bibinfo{publisher}{Springer-Verlag}, \bibinfo{address}{Berlin, Heidelberg}, \bibinfo{pages}{347–361}.
\newblock
\showISBNx{3540423583}


\bibitem[Stray(2023)]%
        {Stray2023}
\bibfield{author}{\bibinfo{person}{Jonathan Stray}.} \bibinfo{year}{2023}\natexlab{}.
\newblock \showarticletitle{{Editorial Values for News Recommenders: Translating Principles to Engineering}}.
\newblock In \bibinfo{booktitle}{\emph{News Quality in the Digital Age} (\bibinfo{edition}{1st} ed.)}. \bibinfo{publisher}{Routledge}, \bibinfo{pages}{15}.
\newblock
\urldef\tempurl%
\url{https://doi.org/10.4324/9781003257998-13}
\showDOI{\tempurl}


\bibitem[Tian et~al\mbox{.}(2021)]%
        {Tian2021}
\bibfield{author}{\bibinfo{person}{Yu Tian}, \bibinfo{person}{Yuhao Yang}, \bibinfo{person}{Xudong Ren}, \bibinfo{person}{Pengfei Wang}, \bibinfo{person}{Fangzhao Wu}, \bibinfo{person}{Qian Wang}, {and} \bibinfo{person}{Chenliang Li}.} \bibinfo{year}{2021}\natexlab{}.
\newblock \showarticletitle{{Joint Knowledge Pruning and Recurrent Graph Convolution for News Recommendation}}.
\newblock \bibinfo{journal}{\emph{SIGIR 2021 - Proceedings of the 44th International ACM SIGIR Conference on Research and Development in Information Retrieval}} (\bibinfo{year}{2021}), \bibinfo{pages}{51--60}.
\newblock
\showISBNx{9781450380379}
\urldef\tempurl%
\url{https://doi.org/10.1145/3404835.3462912}
\showDOI{\tempurl}


\bibitem[Vaswani et~al\mbox{.}(2017)]%
        {transformer_2017}
\bibfield{author}{\bibinfo{person}{Ashish Vaswani}, \bibinfo{person}{Noam Shazeer}, \bibinfo{person}{Niki Parmar}, \bibinfo{person}{Jakob Uszkoreit}, \bibinfo{person}{Llion Jones}, \bibinfo{person}{Aidan~N. Gomez}, \bibinfo{person}{\L{}ukasz Kaiser}, {and} \bibinfo{person}{Illia Polosukhin}.} \bibinfo{year}{2017}\natexlab{}.
\newblock \showarticletitle{Attention is all you need}. In \bibinfo{booktitle}{\emph{Proceedings of the 31st International Conference on Neural Information Processing Systems}} (Long Beach, California, USA). \bibinfo{publisher}{Curran Associates Inc.}, \bibinfo{address}{Red Hook, NY, USA}, \bibinfo{pages}{6000–6010}.
\newblock
\showISBNx{9781510860964}


\bibitem[Vrijenhoek et~al\mbox{.}(2022)]%
        {Vrijenhoek2021_radio}
\bibfield{author}{\bibinfo{person}{Sanne Vrijenhoek}, \bibinfo{person}{Gabriel B\'{e}n\'{e}dict}, \bibinfo{person}{Mateo Gutierrez~Granada}, \bibinfo{person}{Daan Odijk}, {and} \bibinfo{person}{Maarten De~Rijke}.} \bibinfo{year}{2022}\natexlab{}.
\newblock \showarticletitle{RADio – Rank-Aware Divergence Metrics to Measure Normative Diversity in News Recommendations}. In \bibinfo{booktitle}{\emph{Proceedings of the 16th ACM Conference on Recommender Systems}} (Seattle, WA, USA) \emph{(\bibinfo{series}{RecSys '22})}. \bibinfo{publisher}{Association for Computing Machinery}, \bibinfo{address}{New York, NY, USA}, \bibinfo{pages}{208–219}.
\newblock
\showISBNx{9781450392785}
\urldef\tempurl%
\url{https://doi.org/10.1145/3523227.3546780}
\showDOI{\tempurl}


\bibitem[Vrijenhoek et~al\mbox{.}(2021)]%
        {Vrijenhoek2021_mission}
\bibfield{author}{\bibinfo{person}{Sanne Vrijenhoek}, \bibinfo{person}{Mesut Kaya}, \bibinfo{person}{Nadia Metoui}, \bibinfo{person}{Judith M\"{o}ller}, \bibinfo{person}{Daan Odijk}, {and} \bibinfo{person}{Natali Helberger}.} \bibinfo{year}{2021}\natexlab{}.
\newblock \showarticletitle{Recommenders with a Mission: Assessing Diversity in News Recommendations}. In \bibinfo{booktitle}{\emph{Proceedings of the 2021 Conference on Human Information Interaction and Retrieval}} (Canberra ACT, Australia) \emph{(\bibinfo{series}{CHIIR '21})}. \bibinfo{publisher}{Association for Computing Machinery}, \bibinfo{address}{New York, NY, USA}, \bibinfo{pages}{173–183}.
\newblock
\showISBNx{9781450380553}
\urldef\tempurl%
\url{https://doi.org/10.1145/3406522.3446019}
\showDOI{\tempurl}


\bibitem[Vrijenhoek et~al\mbox{.}(2023)]%
        {NORMalize_23_workshop}
\bibfield{author}{\bibinfo{person}{Sanne Vrijenhoek}, \bibinfo{person}{Lien Michiels}, \bibinfo{person}{Johannes Kruse}, \bibinfo{person}{Alain Starke}, \bibinfo{person}{Nava Tintarev}, {and} \bibinfo{person}{Jordi Viader~Guerrero}.} \bibinfo{year}{2023}\natexlab{}.
\newblock \showarticletitle{NORMalize: The First Workshop on Normative Design and Evaluation of Recommender Systems}. In \bibinfo{booktitle}{\emph{Proceedings of the 17th ACM Conference on Recommender Systems}} (Singapore, Singapore) \emph{(\bibinfo{series}{RecSys '23})}. \bibinfo{publisher}{Association for Computing Machinery}, \bibinfo{address}{New York, NY, USA}, \bibinfo{pages}{1252–1254}.
\newblock
\showISBNx{9798400702419}
\urldef\tempurl%
\url{https://doi.org/10.1145/3604915.3608757}
\showDOI{\tempurl}


\bibitem[Wang et~al\mbox{.}(2018)]%
        {Wang2018}
\bibfield{author}{\bibinfo{person}{Zihuan Wang}, \bibinfo{person}{Kyusup Hahn}, \bibinfo{person}{Youngsam Kim}, \bibinfo{person}{Sanghyup Song}, {and} \bibinfo{person}{Jong~Mo Seo}.} \bibinfo{year}{2018}\natexlab{}.
\newblock \showarticletitle{{A news-topic recommender system based on keywords extraction}}.
\newblock \bibinfo{journal}{\emph{Multimedia Tools and Applications}} \bibinfo{volume}{77}, \bibinfo{number}{4} (\bibinfo{year}{2018}), \bibinfo{pages}{4339--4353}.
\newblock
\showISBNx{1104201755130}
\showISSN{15737721}
\urldef\tempurl%
\url{https://doi.org/10.1007/s11042-017-5513-0}
\showDOI{\tempurl}


\bibitem[Wu et~al\mbox{.}(2019a)]%
        {wu2019-npa}
\bibfield{author}{\bibinfo{person}{Chuhan Wu}, \bibinfo{person}{Fangzhao Wu}, \bibinfo{person}{Mingxiao An}, \bibinfo{person}{Jianqiang Huang}, \bibinfo{person}{Yongfeng Huang}, {and} \bibinfo{person}{Xing Xie}.} \bibinfo{year}{2019}\natexlab{a}.
\newblock \showarticletitle{{NPA}: Neural News Recommendation with Personalized Attention}. In \bibinfo{booktitle}{\emph{Proceedings of the 25th ACM SIGKDD International Conference on Knowledge Discovery \& Data Mining}} (Anchorage, AK, USA) \emph{(\bibinfo{series}{KDD '19})}. \bibinfo{publisher}{Association for Computing Machinery}, \bibinfo{address}{New York, NY, USA}, \bibinfo{pages}{2576–2584}.
\newblock
\showISBNx{9781450362016}
\urldef\tempurl%
\url{https://doi.org/10.1145/3292500.3330665}
\showDOI{\tempurl}


\bibitem[Wu et~al\mbox{.}(2019b)]%
        {wu2019-nrms}
\bibfield{author}{\bibinfo{person}{Chuhan Wu}, \bibinfo{person}{Fangzhao Wu}, \bibinfo{person}{Suyu Ge}, \bibinfo{person}{Tao Qi}, \bibinfo{person}{Yongfeng Huang}, {and} \bibinfo{person}{Xing Xie}.} \bibinfo{year}{2019}\natexlab{b}.
\newblock \showarticletitle{Neural News Recommendation with Multi-Head Self-Attention}. In \bibinfo{booktitle}{\emph{Proceedings of the 2019 Conference on Empirical Methods in Natural Language Processing and the 9th International Joint Conference on Natural Language Processing (EMNLP-IJCNLP)}}. \bibinfo{publisher}{Association for Computational Linguistics}, \bibinfo{address}{Hong Kong, China}, \bibinfo{pages}{6389--6394}.
\newblock
\urldef\tempurl%
\url{https://doi.org/10.18653/v1/D19-1671}
\showDOI{\tempurl}


\bibitem[Wu et~al\mbox{.}(2023)]%
        {Wu2023}
\bibfield{author}{\bibinfo{person}{Chuhan Wu}, \bibinfo{person}{Fangzhao Wu}, \bibinfo{person}{Yongfeng Huang}, {and} \bibinfo{person}{Xing Xie}.} \bibinfo{year}{2023}\natexlab{}.
\newblock \showarticletitle{Personalized News Recommendation: Methods and Challenges}.
\newblock \bibinfo{journal}{\emph{ACM Trans. Inf. Syst.}} \bibinfo{volume}{41}, \bibinfo{number}{1}, Article \bibinfo{articleno}{24} (\bibinfo{date}{jan} \bibinfo{year}{2023}), \bibinfo{numpages}{50}~pages.
\newblock
\showISSN{1046-8188}
\urldef\tempurl%
\url{https://doi.org/10.1145/3530257}
\showDOI{\tempurl}


\bibitem[Wu et~al\mbox{.}(2022)]%
        {wu2022big_industry}
\bibfield{author}{\bibinfo{person}{Chuhan Wu}, \bibinfo{person}{Fangzhao Wu}, \bibinfo{person}{Tao Qi}, {and} \bibinfo{person}{Yongfeng Huang}.} \bibinfo{year}{2022}\natexlab{}.
\newblock \showarticletitle{Are Big Recommendation Models Fair to Cold Users?}
\newblock \bibinfo{journal}{\emph{arXiv preprint arXiv:2202.13607}} (\bibinfo{year}{2022}).
\newblock


\bibitem[Wu et~al\mbox{.}(2021)]%
        {Wu2021-newsBert}
\bibfield{author}{\bibinfo{person}{Chuhan Wu}, \bibinfo{person}{Fangzhao Wu}, \bibinfo{person}{Yang Yu}, \bibinfo{person}{Tao Qi}, \bibinfo{person}{Yongfeng Huang}, {and} \bibinfo{person}{Qi Liu}.} \bibinfo{year}{2021}\natexlab{}.
\newblock \showarticletitle{{N}ews{BERT}: Distilling Pre-trained Language Model for Intelligent News Application}. In \bibinfo{booktitle}{\emph{Findings of the Association for Computational Linguistics: EMNLP 2021}}. \bibinfo{publisher}{Association for Computational Linguistics}, \bibinfo{address}{Punta Cana, Dominican Republic}, \bibinfo{pages}{3285--3295}.
\newblock
\urldef\tempurl%
\url{https://doi.org/10.18653/v1/2021.findings-emnlp.280}
\showDOI{\tempurl}


\bibitem[Wu et~al\mbox{.}(2020)]%
        {Wu2020MIND}
\bibfield{author}{\bibinfo{person}{Fangzhao Wu}, \bibinfo{person}{Ying Qiao}, \bibinfo{person}{Jiun-Hung Chen}, \bibinfo{person}{Chuhan Wu}, \bibinfo{person}{Tao Qi}, \bibinfo{person}{Jianxun Lian}, \bibinfo{person}{Danyang Liu}, \bibinfo{person}{Xing Xie}, \bibinfo{person}{Jianfeng Gao}, \bibinfo{person}{Winnie Wu}, {and} \bibinfo{person}{Ming Zhou}.} \bibinfo{year}{2020}\natexlab{}.
\newblock \showarticletitle{{MIND}: A Large-scale Dataset for News Recommendation}. In \bibinfo{booktitle}{\emph{Proceedings of the 58th Annual Meeting of the Association for Computational Linguistics}}. \bibinfo{publisher}{Association for Computational Linguistics}, \bibinfo{address}{Online}, \bibinfo{pages}{3597--3606}.
\newblock
\urldef\tempurl%
\url{https://doi.org/10.18653/v1/2020.acl-main.331}
\showDOI{\tempurl}


\bibitem[Xue et~al\mbox{.}(2024)]%
        {Xue2024}
\bibfield{author}{\bibinfo{person}{Taofeng Xue}, \bibinfo{person}{Zhimin Lin}, \bibinfo{person}{Zijian Zhang}, \bibinfo{person}{Linsen Guo}, \bibinfo{person}{Haoru Chen}, \bibinfo{person}{Mengjiao Bao}, {and} \bibinfo{person}{Peng Yan}.} \bibinfo{year}{2024}\natexlab{}.
\newblock \showarticletitle{Large Scale Hierarchical User Interest Modeling for Click-through Rate Prediction}. In \bibinfo{booktitle}{\emph{Proceedings of the Recommender Systems Challenge 2024}} (Bari, Italy) \emph{(\bibinfo{series}{RecSysChallenge '24})}. \bibinfo{publisher}{Association for Computing Machinery}, \bibinfo{address}{New York, NY, USA}.
\newblock
\showISBNx{979-8-4007-1127-5/24/10}
\urldef\tempurl%
\url{https://doi.org/10.1145/3687151.3687163}
\showDOI{\tempurl}


\bibitem[Zhao et~al\mbox{.}(2024)]%
        {era_llm_2024}
\bibfield{author}{\bibinfo{person}{Zihuai Zhao}, \bibinfo{person}{Wenqi Fan}, \bibinfo{person}{Jiatong Li}, \bibinfo{person}{Yunqing Liu}, \bibinfo{person}{Xiaowei Mei}, \bibinfo{person}{Yiqi Wang}, \bibinfo{person}{Zhen Wen}, \bibinfo{person}{Fei Wang}, \bibinfo{person}{Xiangyu Zhao}, \bibinfo{person}{Jiliang Tang}, {and} \bibinfo{person}{Qing Li}.} \bibinfo{year}{2024}\natexlab{}.
\newblock \showarticletitle{Recommender Systems in the Era of Large Language Models (LLMs)}.
\newblock \bibinfo{journal}{\emph{IEEE Transactions on Knowledge and Data Engineering}} (\bibinfo{year}{2024}), \bibinfo{pages}{1--20}.
\newblock
\urldef\tempurl%
\url{https://doi.org/10.1109/TKDE.2024.3392335}
\showDOI{\tempurl}


\bibitem[Zhou et~al\mbox{.}(2018)]%
        {din_2018}
\bibfield{author}{\bibinfo{person}{Guorui Zhou}, \bibinfo{person}{Xiaoqiang Zhu}, \bibinfo{person}{Chenru Song}, \bibinfo{person}{Ying Fan}, \bibinfo{person}{Han Zhu}, \bibinfo{person}{Xiao Ma}, \bibinfo{person}{Yanghui Yan}, \bibinfo{person}{Junqi Jin}, \bibinfo{person}{Han Li}, {and} \bibinfo{person}{Kun Gai}.} \bibinfo{year}{2018}\natexlab{}.
\newblock \showarticletitle{Deep Interest Network for Click-Through Rate Prediction}. In \bibinfo{booktitle}{\emph{Proceedings of the 24th ACM SIGKDD International Conference on Knowledge Discovery \& Data Mining}} (London, United Kingdom) \emph{(\bibinfo{series}{KDD '18})}. \bibinfo{publisher}{Association for Computing Machinery}, \bibinfo{address}{New York, NY, USA}, \bibinfo{pages}{1059–1068}.
\newblock
\showISBNx{9781450355520}
\urldef\tempurl%
\url{https://doi.org/10.1145/3219819.3219823}
\showDOI{\tempurl}


\bibitem[Zhu et~al\mbox{.}(2022)]%
        {FuxiCTR_22}
\bibfield{author}{\bibinfo{person}{Jieming Zhu}, \bibinfo{person}{Quanyu Dai}, \bibinfo{person}{Liangcai Su}, \bibinfo{person}{Rong Ma}, \bibinfo{person}{Jinyang Liu}, \bibinfo{person}{Guohao Cai}, \bibinfo{person}{Xi Xiao}, {and} \bibinfo{person}{Rui Zhang}.} \bibinfo{year}{2022}\natexlab{}.
\newblock \showarticletitle{{BARS:} Towards Open Benchmarking for Recommender Systems}. In \bibinfo{booktitle}{\emph{{SIGIR} '22: The 45th International {ACM} {SIGIR} Conference on Research and Development in Information Retrieval, Madrid, Spain, July 11 - 15, 2022}}, \bibfield{editor}{\bibinfo{person}{Enrique Amig{\'{o}}}, \bibinfo{person}{Pablo Castells}, \bibinfo{person}{Julio Gonzalo}, \bibinfo{person}{Ben Carterette}, \bibinfo{person}{J.~Shane Culpepper}, {and} \bibinfo{person}{Gabriella Kazai}} (Eds.). \bibinfo{publisher}{{ACM}}, \bibinfo{pages}{2912--2923}.
\newblock
\urldef\tempurl%
\url{https://doi.org/10.1145/3477495.3531723}
\showDOI{\tempurl}


\bibitem[Zhu et~al\mbox{.}(2021)]%
        {FuxiCTR_21}
\bibfield{author}{\bibinfo{person}{Jieming Zhu}, \bibinfo{person}{Jinyang Liu}, \bibinfo{person}{Shuai Yang}, \bibinfo{person}{Qi Zhang}, {and} \bibinfo{person}{Xiuqiang He}.} \bibinfo{year}{2021}\natexlab{}.
\newblock \showarticletitle{Open Benchmarking for Click-Through Rate Prediction}. In \bibinfo{booktitle}{\emph{{CIKM} '21: The 30th {ACM} International Conference on Information and Knowledge Management, Virtual Event, Queensland, Australia, November 1 - 5, 2021}}, \bibfield{editor}{\bibinfo{person}{Gianluca Demartini}, \bibinfo{person}{Guido Zuccon}, \bibinfo{person}{J.~Shane Culpepper}, \bibinfo{person}{Zi~Huang}, {and} \bibinfo{person}{Hanghang Tong}} (Eds.). \bibinfo{publisher}{{ACM}}, \bibinfo{pages}{2759--2769}.
\newblock
\urldef\tempurl%
\url{https://doi.org/10.1145/3459637.3482486}
\showDOI{\tempurl}


\end{thebibliography}

\clearpage
\appendix
\section{EB-NeRD Dataset Overview}
\label{appendix:dataset-overview} 

\begin{table}[htbp]
    \centering
    \captionsetup{justification=raggedright, singlelinecheck=false}
    \caption{
    Detailed description of \textit{articles.parquet}.
    }
    \label{tab:articles_parquet}
    \resizebox{\textwidth}{!}{%
    \begin{tabular}{p{0.1cm}p{2.9cm}p{7.8cm}p{3.2cm}p{2cm}}
    \hline
    \#&Column&Context&Example&dtype \\ \hline \hline
    1  &Article ID&The unique ID of a news article. &8987932 &i32 \\ \hline
    2  &Title  &The article's Danish title. &Se billederne: Zlatans paradis til salg &str \\ \hline
    3  &Subtitle  &The article's Danish subtitle. & Zlatan Ibrahimovic har sat (\dots). &str \\ \hline
    4  &Body  &The article's full Danish text body. & Drømmer du om en eksklusiv (\dots). & str \\ \hline
    5  &Category ID  &	The category ID. &142 & i16	 \\ \hline
    6  &Category string  &The category as a string. &sport &str \\ \hline
    7  &Subcategory IDs  &The subcategory IDs. &[196, 271] & list[i16] \\ \hline
    8  &Premium  &Whether the content is behind a paywall. & False & bool \\ \hline
    9  &Time published	&The time the article was published. The format is \textit{"YYYY/MM/DD HH:MM:SS"}.&	2021-11-15 03:56:56&	datetime[$\mu s$] \\ \hline
    10&	Time modified&	The timestamp for the last modification of the article, e.g., updates as the story evolves or spelling corrections. The format is \textit{"YYYY/MM/DD HH:MM:SS"}.&	2023-06-29 06:38:41&	datetime[$\mu s$] \\ \hline
    11&	Image IDs&	The image IDs used in the article.&	[8988118]	&list[i64] \\ \hline
    12&	Article type&	The type of article, such as a feature, gallery, video, or live blog.&	article\_default&	str \\ \hline
    13&	URL&	The article's URL.&	\href{https://ekstrabladet.dk/sport/fodbold/landsholdsfodbold/se-billederne-zlatans-paradis-til-salg/8987932}{\makecell[l]{https://ekstrabladet.dk/.../\\8987932}} &str \\ \hline
    14&	NER &	The tags retrieved from a proprietary named-entity-recognition model at Ekstra Bladet are based on the concatenated title, abstract, and body.&	['Aftonbladet', 'Sverige', 'Zlatan Ibrahimovic'] &	list[str]  \\ \hline
    15 &	Entities&	The tags retrieved from a proprietary entity-recognition model at Ekstra Bladet are based on the concatenated title, abstract, and body.&	['ORG', 'LOC', 'PER']&	list[str] \\ \hline
    16&	Topics&	The tags retrieved from a proprietary topic-recognition model at Ekstra Bladet are based on the concatenated title, abstract, and body.&	[]&	list[str]\\ \hline
    17&	Total in views&	The total number of times an article has been in view (registered as seen) by users within the first 7 days after it was published. This feature only applies to articles that were published after February 16, 2023.&	null&	i32 \\ \hline
    18&	Total pageviews&	The total number of times an article has been clicked by users within the first 7 days after it was published. This feature only applies to articles that were published after February 16, 2023.&	null&	i32 \\ \hline
    19&	Total read-time&	The accumulated read-time of an article within the first 7 days after it was published. This feature only applies to articles that were published after February 16, 2023.&	null&	f32 \\ \hline
    20&	Sentiment label&	The assigned sentiment label from a proprietary sentiment model at Ekstra Bladet is based on the concatenated title and abstract. The labels are negative, neutral, and positive.&	Neutral&	str \\ \hline
    21&	Sentiment score&	The sentiment score from a proprietary sentiment model at Ekstra Bladet is based on the concatenated title and abstract. The score is the corresponding probability to the \textit{sentiment label}. &	0.5299&	f32
    \\ \hline\hline
    \end{tabular}
    }
\end{table}

\begin{table*}[htbp]
    \centering
    \captionsetup{justification=raggedright, singlelinecheck=false}
    \caption{
    Detailed description of \textit{behaviors.parquet}.
    The training and validation sets have exactly the same format, whereas some features are removed from the test set. These features are \textit{Article ID}, \textit{Next read-time}, \textit{Next Scroll Percentage}, and \textit{Clicked Article IDs}. Furthermore, to include beyond-accuracy computations, we have included 200{,}000 samples. Hence, the test set has an extra called \textit{is beyond-accuracy}.
    }
    \label{tab:behaviors_parquet}
    \resizebox{\textwidth}{!}{%
    \begin{tabular}{p{0.1cm}p{2.9cm}p{7.8cm}p{3.2cm}p{2cm}}
    \hline
    \#&Column&Context&Example&dtype \\ \hline \hline
    1  &Impression ID&The unique ID of an impression. &153 &u32 \\ \hline
    2  &User ID  &The anonymized user ID. &44038 &u32 \\ \hline
    3  &Article ID  &The unique ID of a news article. An empty field means the impression is from the front page. &9650148 &i32 \\ \hline
    4  &Session ID  &A unique ID for a user's browsing session. &1153 &u32 \\ \hline
    5  &In view article IDs  &List of in view article IDs in the impression (news articles that were registered as seen by the user). The order of the IDs have been shuffled. & [9649538, 9649689, \dots, 9649569] &list[i32] \\ \hline
    6  &Clicked article IDs  &List of article IDs clicked in the impression. &[9649689] &list[i32] \\ \hline
    7  &Time  & The impression timestamp. The format is \textit{"YYYY/MM/DD HH:MM:SS"}. &2023-02-25 06:41:40 & datetime[$\mu s$] \\ \hline
    8  &Read-time  &The amount of time, in seconds, a user spends on a given page. &14.0 &f32 \\ \hline
    9  &Scroll percentage  &The percentage of an article that a user scrolls through. & 100.0 & f32 \\ \hline
    10 &Device type  &	The type of device used to access the content, such as desktop (1) mobile (2), tablet (3), or unknown (0). & 1 & i8 \\ \hline
    11 &SSO status  &Indicates whether a user is logged in through Single Sign-On (SSO) authentication. & True &bool \\ \hline
    12 &Subscription status &The user's subscription status indicates whether they are a paid subscriber. Note that the subscription is fixed throughout the period and was set when the dataset was created. &True &bool \\ \hline
    13 &Gender &The user's gender, either male (0) or female (1), as specified in their profile. &null &i8 \\ \hline
    14 &Postcode &The user's postcode, aggregated at the district level as specified in their profile, categorized as metropolitan (0), rural district (1), municipality (2), provincial (3), or big city (4). &2 &i8 \\ \hline
    15 &Age & The user's age, as specified in their profile, categorized into 10-year bins (e.g., 20-29, 30-39, etc.) &50 &i8 \\ \hline
    16 &Next read-time& The time a user spends on the next clicked article, i.e., the article in the \textit{clicked article IDs}. &8.0 &f32 \\ \hline
    17 &Next scroll percentage &The scroll percentage for a user's next article interaction, i.e., the article in \textit{clicked article IDs}. &41.0 &f32 \\ \hline\hline
    \end{tabular}
    }
\end{table*}

\begin{table*}[htbp]
    \centering
    \captionsetup{justification=raggedright, singlelinecheck=false}
    \caption{
    Detailed description of \textit{history.parquet}.
    }
    \label{tab:history_parquet}
    \resizebox{\textwidth}{!}{%
    \begin{tabular}{p{0.1cm}p{2.9cm}p{7.8cm}p{3.2cm}p{2cm}}
    \hline
    \#&Column&Context&Example&dtype \\ \hline \hline
    1  &User ID	&The anonymized user ID.	&44038	&u32\\ \hline
    2  &Article IDs  &The article IDs clicked by the user. &[9618533, … 9646154] & list[i32] \\ \hline
    3   &Timestamps	&The timestamps of when the articles were clicked. The format is \textit{"YYYY/MM/DD HH:MM:SS"}.	& [2023-02-02 16:37:42, \dots, 2023-02-22 18:28:38] &	list[datetime[$\mu s$]] \\ \hline
    4	&Read-times	&The read-times of the clicked articles.	&[425.0, … 12.0]	&list[f32] \\ \hline
    5	&Scroll percentage	&The scroll percentage of the clicked articles.	&[null, \dots 100.0]	&list[f32] 
    \\ \hline\hline
    \end{tabular}
    }
\end{table*}

\end{document}